\documentclass[lettersize,journal]{IEEEtran}

\usepackage{cite}

\usepackage{amsmath,amssymb,amsfonts}
\usepackage{amsthm}

\usepackage{array}
\usepackage{booktabs}
\usepackage{threeparttable}
\usepackage[table]{xcolor}

\usepackage{algorithm}
\usepackage{algorithmic}

\usepackage{graphicx}
\usepackage[caption=false,font=footnotesize]{subfig}

\usepackage{stfloats}   
\usepackage{url}
\usepackage{textcomp}
\usepackage{verbatim}
\usepackage{bbding}
\usepackage{hyperref}
\hypersetup{colorlinks=true, citecolor=blue, linkcolor=blue, urlcolor=blue}

\theoremstyle{plain}
\newtheorem{remark}{Remark}

\begin{document}

\title{
Digital Twin-Assisted Task Offloading and Resource Allocation in ISAC-Enabled Internet of Vehicles}


\author{Shanhao Zhan, Zhang Liu, Lianfen Huang, Shaowei Shen, Ziyang Bai, Zhibin Gao, \\ 
and Dusit Niyato, \textit{Fellow, IEEE}
\thanks{\textit{Shanhao Zhan (shanhao@stu.xmu.edu.cn) is with the Department of Informatics and Communication Engineering, Xiamen University, Fujian 361102, China. Zhang Liu (zhangliu@xmu.edu.cn) is with the Department of Computer Science and Technology, Xiamen University, Fujian 361102, China. Lianfen Huang (lfhuang@xmu.edu.cn) is with the Key Laboratory of Intelligent Manufacturing Equipment and Industrial Internet Technology, Fujian Provincial Universities, and with School of Information Science and Technology, Xiamen University Tan Kah Kee College, and with the Department of Informatics and Communication Engineering, Xiamen University, Fujian 361102, China. Shaowei Shen (shenshaowei@stu.xmu.edu.cn) and Ziyang Bai (23320230157375@stu.xmu.edu.cn) are with the Department of Informatics and Communication Engineering, Xiamen University, Fujian 361102, China. Zhibin Gao (gaozhibin@jmu.edu.cn) is with Navigation Institute, Jimei University, Xiamen 361021, China. D. Niyato (dniyato@ntu.edu.sg) is with the College of Computing and Data Science, Nanyang Technological University, Singapore. (Corresponding author: Zhibin Gao)}}
}



\maketitle

\begin{abstract}
The convergence of the Internet of vehicles (IoV) and 6G networks is driving the evolution of next-generation intelligent transportation systems. However, IoV networks face persistent challenges, including low spectral efficiency in vehicular communications, difficulty in achieving dynamic and adaptive resource optimization, and the need for long-term stability under highly dynamic environments. In this paper, we study the problem of digital twin (DT)-assisted task offloading and resource allocation in integrated sensing and communication (ISAC)-enabled IoV networks. The objective is to minimize the long-term average system cost, defined as a weighted combination of delay and energy consumption, while ensuring queue stability over time. To address this, we employ an ISAC-enabled design and introduce two transmission modes (i.e., raw data transmission (DataT) and instruction transmission (InstrT)). The InstrT mode enables instruction-level transmission, thereby reducing data volume and improving spectral efficiency. We then employ Lyapunov optimization to decompose the long-term stochastic problem into per-slot deterministic problems, ensuring long-term queue stability. Building upon this, we propose a Lyapunov-driven DT-enhanced multi-agent proximal policy optimization (Ly-DTMPPO) algorithm, which leverages DT for global state awareness and intelligent decision-making within a centralized training and decentralized execution (CTDE) architecture. Extensive simulations verify that Ly-DTMPPO achieves superior performance compared with existing benchmarks.
\end{abstract}         

\begin{IEEEkeywords}
Integrated sensing and communications, digital twin, resource allocation, task offloading, Lyapunov optimization.
\end{IEEEkeywords}

\section{Introduction}\label{sec1}
\subsection{Background Research}\label{subsec1} 
\IEEEPARstart{W}{ith} the commercial deployment of 5G and the ongoing research toward 6G, the Internet of vehicles (IoV) has emerged as a cornerstone technology for intelligent transportation and autonomous driving. However, IoV faces pressing demands for low latency, highly reliable communications, accurate environmental sensing, and intelligent computing \cite{9817611}. Many IoV applications are computation-intensive and latency-sensitive (e.g., autonomous driving, high-definition video analytics, and cooperative vehicular perception), where multi-sensor data from on-board cameras, LiDAR, and millimeter-wave radars must be processed in real time \cite{10375781}. The deep neural networks (DNNs) that underpin these applications are typically large-scale; for example, ResNet-152 requires roughly 22.6 billion operations for a single image classification, and processing high-definition video sequences can exceed 30 trillion operations \cite{10736570}. Relying solely on on-board computing for such workloads leads to prohibitive inference delays, undermining safety and user experience.

To mitigate this bottleneck, vehicular edge computing (VEC) has been proposed as an effective task-offloading paradigm \cite{10.1109/TITS.2024.3388422,10.1109/TITS.2024.3351635}. Unlike centralized cloud computing, VEC pushes computation and storage to the network edge. Thus, tasks can be partitioned and jointly executed across roadside units (RSUs) via vehicle-to-infrastructure (V2I) links and nearby vehicles via vehicle-to-vehicle (V2V) links. This edge-proximal architecture significantly reduces end-to-end latency and alleviates backhaul congestion, enabling timely execution of computation-intensive and latency-sensitive IoV applications  \cite{10.1109/TITS.2024.3349546}.

Furthermore, the limited spectrum resource has become a major bottleneck for IoV, which must support ultra-reliable communications and high-throughput data exchange for real-time services. For instance, modern autonomous vehicles rely on a large number of onboard sensors that can produce over 8 Gbps of data \cite{11175176}. Integrated sensing and communication (ISAC) has been recognized as a key enabler for 6G to address this limitation, as it performs both communications and environment sensing within the same spectrum \cite{9737357}. Unlike the traditional paradigm of “sensing–communication separation,” ISAC enables nodes to perform both communication and sensing, which not only improves spectrum efficiency but also expands the effective communication bandwidth, thereby increasing data transmission rates \cite{10812728}.

However, ISAC-assisted IoV still suffers from reactive decision-making due to the lack of prediction capability. While ISAC can sense the current communication and environmental conditions, it cannot foresee future mobility and workload variations. As a result, task offloading and resource scheduling remain prone to inefficiency under high mobility. Digital twin (DT), as a pivotal tool for cyber–physical integration, addresses this gap by offering global visibility and predictive intelligence \cite{10423175}. By constructing high-fidelity virtual replicas of vehicles, RSUs, and network resources, DT enables real-time synchronization of physical states and supports predictive modeling of task dynamics. For instance, DT can predict vehicle trajectories and task load distributions, thereby facilitating proactive optimization of offloading and resource allocation \cite{10234400}. Consequently, combining DT with ISAC provides a more comprehensive network optimization framework for next-generation intelligent transportation systems.

\subsection{Motivation and Main Challenges}\label{subsec2}
Although integrating DT and ISAC technologies into the IoV offers significant potential, several fundamental challenges remain unresolved.

\begin{itemize}
    \item \textbf{\textit{First, effectively exploiting ISAC to enhance communication in IoV is a critical challenge.}} The tight coupling between sensing and communication creates nontrivial trade-offs between throughput and perception accuracy, making resource allocation highly state-dependent and more difficult to stabilize \cite{s25030723}. These issues are further exacerbated in dense, high-mobility scenarios, where simultaneous demands for high-rate transmission and precise sensing significantly increase scheduling uncertainty \cite{10115012}. To address this, we leverage ISAC’s dual functionality to support two transmission modes: raw data transmission (DataT) and instruction transmission (InstrT), where RSUs reconstruct sensory data using their own perception. This mechanism effectively reduces transmission overhead and latency while ensuring task execution accuracy.
    
    \item \textbf{\textit{Second, minimizing latency and energy consumption while ensuring long-term stability poses a fundamental challenge.}} In many practical deployments, both vehicles and RSUs are subject to limited energy budgets \cite{liu2021vehicular,10580985}. Under highly dynamic mobility and time-varying task arrivals, excessive short-term energy consumption can easily lead to queue backlogs and degrade service reliability. Without foresight of future system dynamics, making consistent resource allocation decisions across time slots becomes extremely difficult, especially when simultaneously pursuing low delay and energy efficiency \cite{9895362}. To address this challenge, we employ a Lyapunov optimization framework that decomposes the long-term stochastic control problem into tractable per-slot decisions, dynamically balancing delay reduction and energy consumption while guaranteeing queue stability over time. This design ensures that vehicles and RSUs can sustain continuous operation without sacrificing responsiveness in highly dynamic IoV environments.
    
    \item \textbf{\textit{Third, achieving efficient and adaptive task offloading in highly dynamic IoV environments remains a major challenge.}} Vehicular tasks are heterogeneous and delay-sensitive, while high mobility leads to frequent topology variations and fluctuating workloads \cite{10423175,10365525}. Traditional optimization methods (e.g., convex optimization and heuristic algorithms) struggle to make real-time decisions in such environments. The deep reinforcement learning (DRL) provides adaptability through environment interaction mechanisms, but agents relying solely on local observations may produce unstable or short-sighted actions under partial observability. To tackle this challenge, we integrate DT into the DRL framework under the centralized training and decentralized execution (CTDE) paradigm, where synchronized global states are provided during training. This improves policy awareness of system dynamics and enhances decision robustness in rapidly changing scenarios (as validated in Sec.~\ref{sec7}).
\end{itemize}

\subsection{Summary of Contributions}\label{subsec3}
In this paper, motivated by the above challenges, we investigate DT-assisted task offloading and resource allocation in lSAC-enabled loV. Our objective is to minimize the long-term system cost, defined as a weighted sum of the average delay and energy consumption, while ensuring queue stability in highly dynamic vehicular environments. The main contributions of this work are summarized as follows:

\begin{itemize}
\item \textbf{\textit{Framework:}}
We formulate a long-term joint task offloading and resource allocation problem in ISAC-enabled IoV, where vehicles and RSUs adopt dual transmission modes (i.e., DataT and InstrT), with the latter reducing data volume and improving spectral efficiency. We model the problem as a mixed-integer nonlinear programming (MINLP) formulation, which is known to be NP-hard. It is particularly challenging to solve under high vehicle mobility and time-varying task arrivals.

\item \textbf{\textit{Solution:}}
To address this challenge, we first employ the Lyapunov optimization technique to transform the original long-term coupled objective into a sequence of tractable per-slot subproblems, ensuring queue stability and sustainable energy consumption. Building on this transformation, we propose a Lyapunov-driven DT-enhanced multi-agent proximal policy optimization (Ly-DTMPPO) algorithm under the CTDE paradigm. The algorithm integrates DT-enhanced global visibility into on-policy reinforcement learning and embeds Lyapunov stability awareness into the reward design, achieving faster convergence, superior stability, and improved adaptability compared with baselines.

\item \textbf{\textit{Validation:}}
We conduct extensive simulations to validate the proposed framework. The results show that Ly-DTMPPO achieves faster convergence, better stability, and greater reductions in both delay and energy consumption compared with Lyapunov-driven multi-agent baselines, thereby attaining the lowest overall system cost. Moreover, the comparison with its DT-absent counterpart confirms that the proposed DT module significantly enhances decision stability and improves the effectiveness of computation offloading and resource scheduling in dynamic vehicular environments.
\end{itemize}
\subsection{Paper Organization}\label{subsec4}
The remainder of this paper is organized as follows. Sec. \ref{sec2} reviews related works on DT-assisted ISAC in IoV and Lyapunov-based optimization approaches. Sec. \ref{sec3} introduces the system model, including DT-assisted vehicular scenarios and ISAC-enabled communication, computing, and queue models. Sec. \ref{sec4} formulates the joint optimization problem with latency and energy constraints. Sec. \ref{sec5} presents the Lyapunov-based problem decomposition method. Sec. \ref{sec6} develops the proposed Ly-DTMPPO algorithm under a multi-agent reinforcement learning framework. Sec. \ref{sec7} evaluates system performance through simulations and comparative analysis. Finally, Sec. \ref{sec8} concludes the paper and discusses future research directions.

\section{Related Work}\label{sec2}
Henceforth, we review the most relevant studies and identify the research gaps that motivate the present work.

\subsection{ISAC-Enabled Vehicular Networks}
As a fundamental technology for 6G, ISAC enables simultaneous communication and sensing over shared spectrum. The authors in \cite{10606449} developed an ISAC-enabled V2X MEC framework to reduce long-term service delay via joint offloading and resource allocation. A three-tier ISCC architecture utilizing MEC and cloud servers was proposed in \cite{10795233} to improve latency and energy performance for sensing tasks. In addition, a data-fusion-based ISAC approach was designed in \cite{10007643} to optimize long-term delay–energy trade-offs, and \cite{10005825} proposed an ISAC-driven association strategy to expand sensing range and reduce communication latency in ad hoc networks.

Despite these advancements, existing studies largely concentrate on physical-layer signal processing or short-term trade-off optimization. The interaction between ISAC and higher-layer decision-making, such as cross-layer resource allocation, task scheduling, and long-term system stability, remains insufficiently addressed.

\subsection{DT-Assisted ISAC in IoV}
DT has emerged as a key technology for the IoV, offering global visibility and predictive modeling of network states. The authors in \cite{10234400} proposed a DT-driven edge–end scheduling scheme that improves heterogeneous task completion via accurate resource estimation. The work in \cite{11023535} examined the trade-off between maintaining DT models and executing tasks, highlighting DT’s role in balancing scarce computing resources. A collaborative driving architecture leveraging DT for real-time vehicular interaction was presented in \cite{10319104}. Additionally, Li et al. \cite{10190734} investigated DT-assisted integrated sensing–communication–computation, where DT predictions assist in offloading decisions to balance sensing accuracy and computational efficiency.

However, most existing DT-assisted IoV studies focus primarily on architecture design or predictive scheduling, while only a limited number explicitly address the tight coupling among sensing, communication, and computing resources in ISAC-enabled environments \cite{9982429}. The integration of DT and ISAC for stable long-term resource allocation under high mobility remains largely underexplored.

\subsection{Resource Allocation for ISAC-Enabled IoV }
Efficient resource allocation is essential for ISAC-enabled IoV, where spectrum, computing, and power resources must be coordinated to support diverse and latency-critical tasks. The authors in \cite{10.1145/3636534.3698232} optimized SDMA-based scheduling to maximize throughput in ISAC-enabled IoV. Chen et al. \cite{chen2024integratedsensingcommunicationcomputing} introduced the ISCC paradigm with TSFC resource decoupling managed through A2GNN. The work in \cite{10606449} jointly optimized offloading and resource allocation to reduce long-term delay, and \cite{10007643} leveraged Lyapunov optimization to balance queue latency and energy in cooperative perception.

Although effective, these studies largely emphasize instantaneous or simulation-driven optimization and often simplify the sensing–communication coupling, leaving long-term queue and energy stability underexplored. This gap motivates stability-aware, cross-layer resource allocation tailored to DT-enabled ISAC-aided IoV.

\subsection{Lyapunov Optimization-Based Approaches on IoV}
Lyapunov optimization is widely used in IoV to guarantee queue stability while optimizing long-term metrics. The authors in \cite{10283781} applied it to ISAC networks for joint bandwidth and power allocation to reduce latency and improve sensing. In the MEC context, \cite{10115012} adopted a Lyapunov-based multi-agent approach for task and resource coordination, while \cite{9895362} integrated Lyapunov optimization with GCN-based RL to manage interdependent subtasks. Furthermore, \cite{10638833} presented LySAC, combining Lyapunov with soft actor–critic to jointly optimize delay and energy.

However, these studies typically rely on simplified assumptions or address only isolated communication–computation trade-offs. The coupling among sensing, communication, and computation, along with the need for predictive global visibility, remains largely unexplored. This motivates our deeper investigation into DT-assisted ISAC-enabled IoV systems using Lyapunov optimization to achieve long-term stability.

\section{System Model}\label{sec3}
In this section, we present the DT-assisted ISAC-enabled IoV system model. We first describe the DT-assisted vehicular network architecture and ISAC-enabled communication and computing scenarios, followed by the modeling of task offloading, resource allocation, and queue dynamics. They together lay the foundation for the Lyapunov-based optimization and learning framework developed later.
\begin{figure*}[htbp]
\centering
\includegraphics[width=7in]{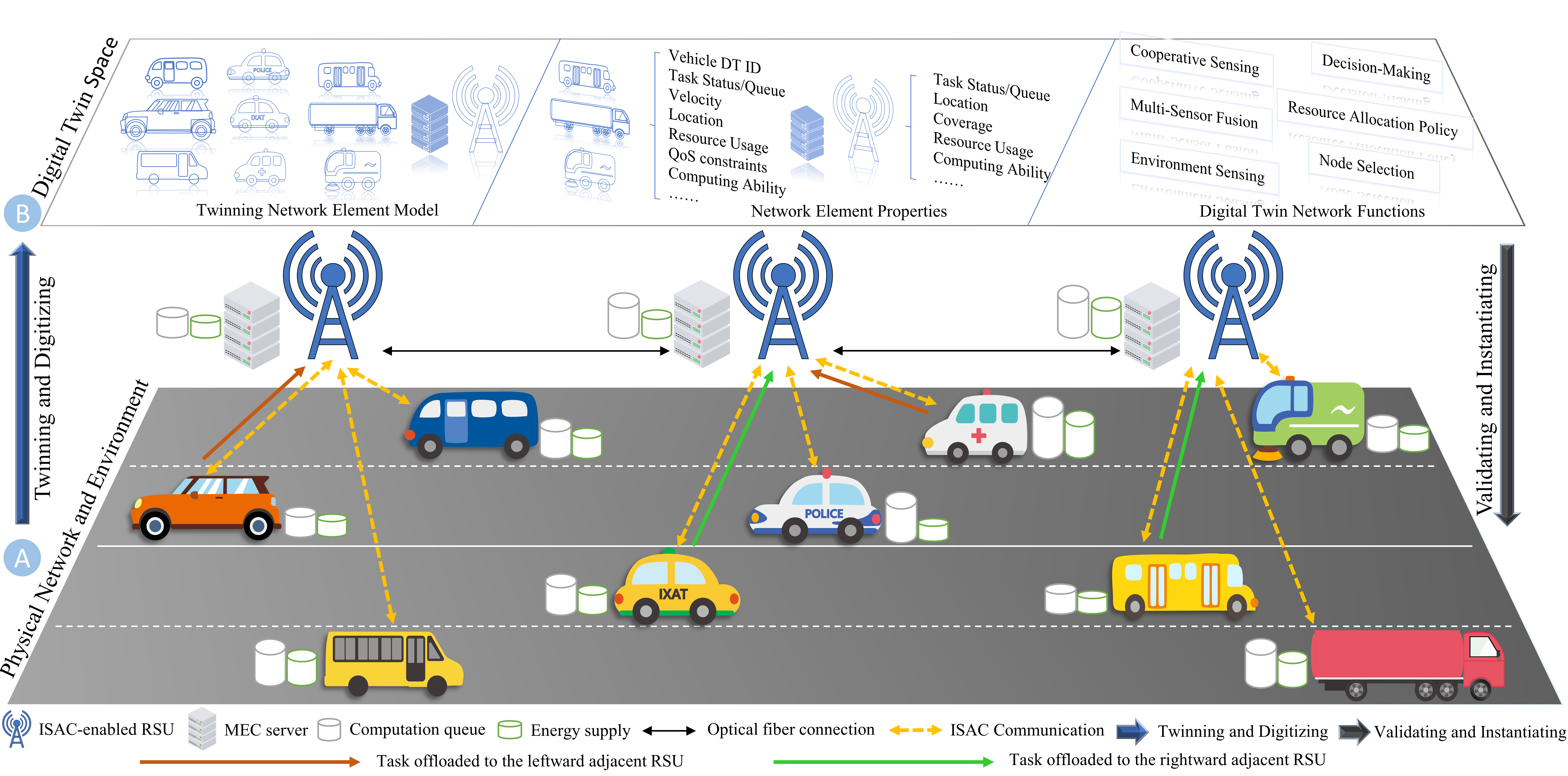}
\caption{DT-Assisted ISAC vehicular network scenario. Part A illustrates the physical layer, where ISAC-enabled vehicles and RSUs cooperate in sensing, communication, and computation. Therein, ISAC communication is performed between the vehicle and the RSU. Part B presents the DT layer, which mirrors the physical environment through synchronized digital replicas, supporting predictive control and proactive resource optimization.}
\label{fig_1}
\end{figure*}

\subsection{DT-Assisted IoV Scenarios}\label{subsec31}
As shown in Fig. \ref{fig_1}, we introduce a vehicular network scenario that integrates communication, sensing, and computation, and consists of several RSUs deployed along a two-way straight city road segment, where at regular intervals (dependent on the RSU range). These RSUs, equipped with ISAC capabilities and integrated with the MEC server \cite{10638833}, offering computational capacities surpassing those of mobile vehicles, and sensors (e.g. high-definition cameras, LiDAR, and millimeter-wave radar) \cite{8907851} to sense the information of environment in real time. Adjacent RSUs, interconnected through optical fibers, have the capability to create a DT space to enable a global situational awareness of communication, sensing, and computation states, thereby facilitating accurate and informed decision-making. In our paper, to simultaneously optimize both time-domain and frequency-domain resource allocation, the orthogonal frequency division multiple access (OFDMA) technique is employed to enable concurrent communication between CVs and RSUs. As a result, the interference caused by signal overlapping can be neglected in the subsequent analysis \cite{10736570,10115012}. Moreover, we adopt a time-division duplexing (TDD) mode for the uploading and feedback process.

We adopt a discrete time-slot computational model where the total duration is divided into $\mathrm{T}$  equal time slots, and the duration of each time slot is denoted as $\tau$. The set of time solts are represented as $\mathcal{T}=\{1, \ldots, \mathrm{T} \}$. The set of RSUs can be described as $\mathbb{R}=\{\mathcal{R}_1, \ldots, \mathcal{R}_K\}$ ($k \in \{1, \ldots, K\}$). Some vehicles carry a computation-intensive and latency-intensive task that takes environmental data as input. These vehicles, regarded as Client Vehicles (CVs), can be described as $\mathbb{C}=\{\mathcal{C}_1, \ldots, \mathcal{C} _I\}$ ($i \in \{1, \ldots, I\}$). The details of the vehicle and RSU DT are as follows:

\textit{1) CV-Twin Body:} CV digital twins include task data volume $G_{i}^{\mathrm{TasD}}$ and computation instruction data volume $G_{i}^{\mathrm{InsD}}$, computation capability $C_{CV_i}^{\mathrm{cpu}}$, vehicle DT ID $DT-Number_{CVs}$, coordinates $L_{i}^{\mathrm{cv}}$, the speed of vehicle $v_{i}^{\mathrm{cv}}$, and the maximum tolerable latency $T_{i}^{\mathrm{max}}$, where $\mathcal{C}_i\in \mathbb{C}$.

\textit{2) RSU-Twin Body:} RSU digital twins include RSU DT ID $DT-Number_{\mathrm{RSU}}$, coordinates $L_{k}^{\mathrm{rsu}}$, coverage area (i.e. radius) $Range_{k}^{\mathrm{rsu}}$, computing resources $F_k^{\mathrm{rsu,max}}$, where $\mathcal{R}_k\in \mathbb{R}$.

Unlike conventional IoV models where vehicles can only access local information, the DT space in our framework provides a virtualized global view of the system. Specifically, the DT synchronizes the physical states of vehicles and RSUs in real time and maintains virtual queues that mirror the task backlog and resource utilization in RSUs. These queues maintained by the DT are not directly observable by vehicles in practice, but become available to the intelligent decision-making agents through the DT space. This enables predictive modeling of mobility and task dynamics, allowing the reinforcement learning algorithm to exploit global context information for more accurate task offloading and resource allocation.

Accordingly, the DT not only records static parameters (e.g., location and computing capacity) but also functions as a dynamic estimator and predictor, which bridges the gap between locally observable states and globally optimal scheduling strategies. In our practical implementation, the DT further provides the actor network in the CTDE framework with a global perspective, enabling agents to make more informed and coordinated decisions. The effectiveness of this DT-assisted design is validated in our experiments, as detailed in Sec. \ref{subsec7.3}. For ease of reference, Table \ref{tab:symbols} summarizes the key notations used in the system model.

\begin{table}[htbp]
\caption{List of Key Notations}
\centering
\rowcolors{2}{gray!20}{white}
\begin{tabular}{|c|p{6.2cm}|}
\hline
\textbf{Symbol} & \textbf{Description} \\
\hline
$\textbf{b}(t)$ & Bandwidth allocation decision \\
$\mathbb{C}$ & The vehicle set \\
$C_{CV_i}^{\mathrm{cpu}}$ & Local computing capacity of vehicle $\mathcal{C}_i$ \\
$\complement(t)$ & The system-level cost function \\
$d_{i,k}^{\mathrm{veh\to rsu}}(t)$ & The Euclidean distance between CV $\mathcal{C}_i$ and RSU $\mathcal{R}_k$ \\
$D_{i}^{\mathrm{loc}}(t)$  & The local processing latency in CV $\mathcal{C}_i$ \\
$D_{i,k}^{\mathrm{rsu}}(t)$ & The edge processing latency in RSU $\mathcal{R}_k$ \\
$D_{i}^{\mathrm{queue,n}}(t)$ & The queueing delay \\
$E_{i}^{\mathrm{loc}}(t)$  & The local processing energy consumption in CV $\mathcal{C}_i$ \\
$E_{i,k}^{\mathrm{DataT-up}}(t)$ & The energy consumption during task uploading in DataT mode \\ 
$E_{i,k}^{\mathrm{co-tra}}(t)$ & The energy consumption of the RSU execute coordinate transformation\\
$E_{i}^{\mathrm{task}}(t)$ & The total energy consumed by all nodes \\ 
$F_{RSU_k}^{\mathrm{cpu}}(t)$ & Computational resource of the RSU allocated to its offloading task at time solt $t$ \\
$G_{i}^{\mathrm{TasD}}$ & Data volume of the task \\
$G_{i}^{\mathrm{InsD}}$ & Data volume of computation instruction \\
$I_{c}^{\mathrm{task}}$ & The number of CPU cycles required to process a bit \\
$\ell(t)$ & Task assignment decision \\
$Q_{i}^{\mathrm{loc}}(t)$ & The task queue length of vehicle $\mathcal{C}_i$ at time slot $t$ \\
$Q_{k}^{\mathrm{rsu}}(t)$ & The task queue length of the RSU $\mathcal{R}_k$ at time slot $t$ \\
$\mathbb{R}$ & The RSU set \\
$R_{i,k}^{\mathrm{veh\to rsu}}(t)$ & The V2I transmission rate from CV $\mathcal{C}_i$ to RSU $\mathcal{R}_k$\\
$\mathcal{T}$ & Index set of time slots \\
$T_i^{\max}$ & The maximum tolerated delay of the task completion\\
$T_{i}^{\mathrm{task}}(t)$ & The overall latency of all tasks executed in paralle \\
$V_{i}(t)$ & The virtual energy queue at time slot $t$ \\
$\Psi_{i}^{\mathrm{CV}}$ & Task of vehicle $\mathcal{C}_i$ at time slot $t$ \\
$\Theta$ & Communication gains \\
$\eta_{i}^{\mathrm{rsu}}(t)$ & The mode selector \\
\hline
\end{tabular}
\label{tab:symbols}
\end{table}

\subsection{ISAC-Enabled IoV Computing Scenarios}\label{subsec32}
In Sec. \ref{subsec31}, with the continuous updates, this twin body updates in real-time to reflect the state of the physical object \cite{9982429}. Therefore, RSU-twin bodies and vehicles-twin bodies in this paper can be regarded as ISAC nodes. Both RSUs and vehicles that are equipped with sensors, their have the ability to get environmental information. Therefore, twin bodies of neighboring RSUs adjacent to a CV may collect similar environmental sensing data. In this case, CVs and RSUs can jointly provide both communication and sensing functions, which further enable diverse computing scenarios in ISAC-assisted IoV systems. This redundancy enables cooperative mechanisms where CVs do not necessarily need to upload raw data for every task.

Specifically, we introduce two available transmission modes for CVs to offload their tasks to RSU-twin bodies:
\begin{itemize}
    \item \textbf{\textit{DataT Mode:}} the CV directly uploads the full raw sensory data $G_{i}^{\mathrm{TasD}}(t)$ (e.g., high-definition video and LiDAR point clouds) to the RSU. The offloading time depends on the transmission rate and the size of the raw task data. After the RSU finishes computation, the output results are sent back to the CV, where the feedback delay depends on the size of the result data and the transmission rate.
    
    \item \textbf{\textit{InstrT Mode:}} instead of transmitting massive raw data, the CV only uploads computation instructions with significantly smaller size $G_{i}^{\mathrm{InsD}}(t)$. Since the RSU itself collects similar environmental data from its own sensors, it can reconstruct the task input using its local perception data after receiving the instructions \cite{9982429}. However, due to the different perspectives between the CV and RSU, the raw environmental data need to be aligned through coordinate transformation. This process can be realized by matrix operations using the coordinate information of the CV contained in the computation instructions, as suggested in \cite{9982429,9439524}. The corresponding computational intensity for this preprocessing is denoted by $I_{c}^{\mathrm{co-task}}$. Once aligned, the RSU executes the task using its own perception data, and returns the result to the CV.
\end{itemize}

By leveraging ISAC’s dual capability, the InstrT mode can significantly reduce transmission overhead and latency compared to DataT, while still ensuring accurate task execution. This ISAC-assisted collaborative mechanism allows tasks to be executed in parallel across RSU-twin bodies, thereby improving both resource utilization and service efficiency in dynamic vehicular environments. In this paper, we focus on scenarios where CVs generate tasks, while RSUs serve as the primary service nodes for task execution. The choice between DataT and InstrT modes becomes a crucial part of the task offloading strategy, and will be jointly optimized with resource allocation in the proposed Lyapunov-driven framework.

\subsection{Sensing-Enhanced Communication Model}\label{subsec33}
In this paper, we consider V2I communication models. Moreover, similar to \cite{10638833}, we consider the collaborative benefits between communication and sensing functions (i.e. sensing-enhanced communication model). Let the enhancement factor $\Theta=R_{sen-enhance}/R_{normal}$ represent the ratio of the communication rate under sensing-enhanced to the normal communication rate. 

\textit{V2I Communication:} Let $R_{i,k}^{\mathrm{veh\to rsu}}(t)$ be the data transmission rate from CV $\mathcal{C}_i$ to the RSU $\mathcal{R}_k$ at time slot $t$:
\begin{align}
\label{V2I_Communication_model}
R_{i,k}^{\mathrm{veh\to rsu}}(t) &= \Theta b_{i,k}^{\mathrm{veh\to rsu}}(t) \mathcal{B} \log_{2} \notag \\
& \Bigg(1 + \frac{P_i \left| h_{i,k}^{\mathrm{veh\to rsu}}(t) \right|^{2} g_{i,k}^{\mathrm{veh\to rsu}}(t)}{\sigma_0^{2}} \Bigg),
\end{align}
where $b_{i,k}^{\mathrm{veh\to rsu}}(t)\mathcal{B}$ denotes the communication bandwidth between the CV $\mathcal{C}_i$ and RSU $\mathcal{R}_k$, $b_{i,k}^{\mathrm{veh\to rsu}}(t)\in [0,1]$ is the bandwidth allocation ratio from CV $\mathcal{C}_i$ to RSU $\mathcal{R}_k$ at time slot $t$, $P_i$ is the transmit power of CV $\mathcal{C}_i$, and $\sigma_0^{2}$ is the additive white gaussian noise (AWGN). Additionally, $h_{i,k}^{\mathrm{veh\to rsu}}(t) \sim \mathcal{CN}(0,1) $ is the small-scale fading component between CV $\mathcal{C}_i$ and RSU $\mathcal{R}_k$ at time slot $t$. In addition, $g_{i,k}^{\mathrm{veh\to rsu}}(t)$ is the path loss of the large-scale fading component between CV $\mathcal{C}_i$ and RSU $\mathcal{R}_k$, calculated as $-38.4-21.0log_{10}(\textit{dis}_{i,k}^{\mathrm{veh\to rsu}}(t))$ \cite{8944302}, where $\textit{dis}_{i,k}^{\mathrm{veh\to rsu}}(t)$ represents the time-varying distance between CV $\mathcal{C}_i$ and RSU $\mathcal{R}_k$ at time slot $t$. 

In addition, due to the high mobility of vehicles, we model the real-time 3D Euclidean distance between vehicle and RSU, to capture the spatial and temporal variation in V2I distance. The Euclidean distance between CV $\mathcal{C}_i$ and RSU $\mathcal{R}_k$ at time slot $t$ is given by
\begin{equation}
\label{V2I_distance_model}
d_{i,k}^{\mathrm{veh\to rsu}}(t) = \sqrt{\left| x_{k}^{\mathrm{rsu}} - x_{i}(t)  \right|^{2} + (y_{i}^{\mathrm{cv}}-l_{k})^{2} + H_{k}^{2}},
\end{equation}
where $x_{i}(t) =x_{i}+ v_{i}^{\mathrm{cv}}t$ is the CV's longitudinal position. $y_{i}^{\mathrm{cv}}$ denotes the lateral position determined by the lane it occupies. The coordinate of RSU $\mathcal{R}_k$ is described as $L_{k}^{\mathrm{rsu}}$ (i.e. a tuple $(x_{k}^{\mathrm{rsu}}, l_{k}, H_{k})$), where $x_{k}^{\mathrm{rsu}}$, $l_{k}$, and $H_{k}$ represent the RSU's longitudinal position, RSU' lateral offset from the road centerline, and the height of RSU, respectively. Similar to \cite{9982429}, we can ignore the relative moving distance between CV $\mathcal{C}_i$ and RSU $\mathcal{R}_k$. Therefore, the Euclidean distance between CV $\mathcal{C}_i$ and RSU $\mathcal{R}_k$ can be updated by
\begin{equation}
\label{V2I_distance_model_real}
d_{i,k}^{\mathrm{veh\to rsu}}(t) = \sqrt{\left| x_{k}^{\mathrm{rsu}}  \right|^{2} + (y_{i}^{\mathrm{cv}}-l_{k})^{2} + H_{k}^{2}}.
\end{equation}


\subsection{Computing Model}\label{subsec34}

\begin{figure}[htbp]
\centering
\includegraphics[width=3.3in]{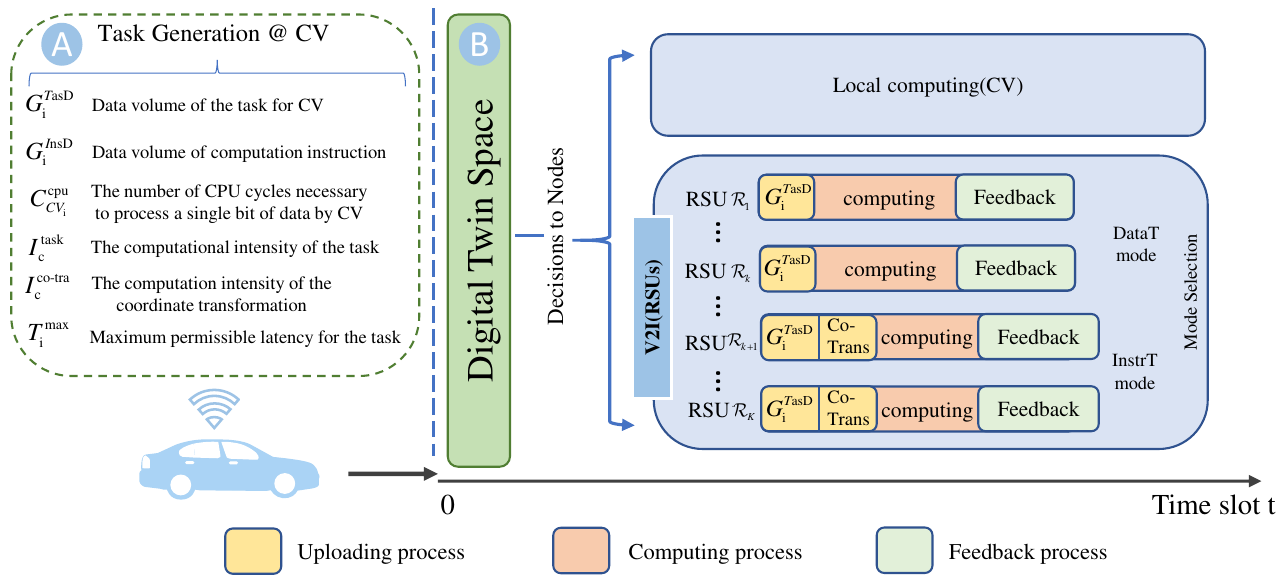}
\caption{Total task completion latency. Here, “Co-Trans” stands for coordinate transformation. Part A illustrates task generation at the CV, where data volume and computational attributes are defined. Part B shows the DT space that decides task-offloading ratios between local and RSU nodes to achieve low-latency cooperative computation.} 
\label{fig_2}
\end{figure}

As shown in Fig. \ref{fig_2}, at the beginning of each time slot $t$, CVs generate a task, i.e.
\begin{equation}
\label{CVs_TASK}
\Psi_{i}^{\mathrm{CV}}=\{ G_{i}^{\mathrm{TasD}}(t), \!G_{i}^{\mathrm{InsD}}(t), \!C_{CV_i}^{\mathrm{cpu}}, \!I_{c}^{\mathrm{task}}, \!I_{c}^{\mathrm{co-task}}, \!T_{i}^{\mathrm{max}} \}, 
\end{equation}
where $G_{i}^{\mathrm{TasD}}(t)$ indicates the data volume of the task for CV $\mathcal{C}_i$ at the time slot $t$. $G_{i}^{\mathrm{InsD}}(t)$ represents the data volume of computation instruction. $C_{CV_i}^{\mathrm{cpu}}$ (in cycles/s) denotes the number of CPU cycles per second that the CV $\mathcal{C}_i$ can execute. $I_{c}^{\mathrm{task}}$ (in cycles/bit) denotes the computational intensity of a task (i.e., the number of CPU cycles necessary to process a single bit of task data). $I_{c}^{\mathrm{co-task}}$ (in cycles/bit) is the computation intensity of the coordinate transformation. $T_{i}^{\mathrm{max}}$ represents the maximum tolerable latency for the task completion. 

According to Sec. \ref{subsec32},  compared with data volume of the task, the task description and corresponding decisions are very small. As a result, these two transmission time can be negligible \cite{9982429}. As shown in Fig. \ref{fig_2}, tasks generated by CV will be computed in parallel on the local and RSU.

\textit{1) Local Processing:} At time slot $t$, when a subtask is executed locally, the local processing latency is the local computation time. The local processing latency $D_{i}^{\mathrm{loc}} (t)$ and the energy consumption of locally executed tasks $E_{i}^{\mathrm{loc}}(t)$, are given by
\begin{align}
\label{local_processing_latency_and_energy_consumption}
& D_{i}^{\mathrm{loc}}(t)=\frac{\ell_{i}^{\mathrm{loc}}(t) G_{i}^{\mathrm{TasD}}(t) I_{c}^{\mathrm{task}}}{C_{CV_i}^{\mathrm{cpu}}}, \\
& E_{i}^{\mathrm{loc}}(t)= \kappa_{cv} D_{i}^{\mathrm{loc}} (t) (C_{CV_i}^{\mathrm{cpu}})^{3}, 
\end{align}
where $\ell_{i}^{\mathrm{loc}}(t)$ represents the ratio of tasks assigned to local. $\kappa_{cv}$ is the energy consumption parameter for vehicle computing, which is a hardware-dependent constant.

\textit{2) Edge Processing at RSU:} When the task is executed at RSU, the CV needs to consider whether to select DataT mode or InstrT mode. Additionally, the data volume of the task or computing instructions will be transmitted to the RSU. After the RSU completes the computation task, the calculation results will be returned to the CV. According to \cite{9982429,2944926,s21082628}, the size of the output result is much smaller than the input data, so we ignore the feedback delay. Therefore, the RSU's task processing latency will further consist of uploading time and processing time, which are formalized below.

\begin{itemize}
    \item \textit{Uploading Time:} If the CV has selected the DataT mode, the uploading time can be calculated by
    \begin{align}
    \label{RSU_processing_latency_DataT_Mode_uploading}
    & D_{k,DataT}^{\mathrm{rsu}} (t)=\frac{\ell_{i,k}^{\mathrm{rsu}}(t) G_{i}^{\mathrm{TasD}}(t)}{R_{i,k}^{\mathrm{veh\to rsu}}(t)},
    \end{align}
     
    where $\ell_{i,k}^{\mathrm{rsu}}(t)$ is the ratio of tasks assigned from CV $\mathcal{C}_i$ to RSU $\mathcal{R}_k$ at time slot $t$.

    Then, $E_{i,k}^{\mathrm{DataT-up}}$ is the energy consumption during task uploading of DataT mode, which is given by
    \begin{equation}
    \label{RSU_processing_energy_consumption_DataT_Mode_uploading}
    E_{i,k}^{\mathrm{DataT-up}}(t) = D_{k,DataT}^{\mathrm{rsu}}(t) P_i(t).
    \end{equation}
    
    The uploading time for InstrT mode is calculated by
    \begin{align}
    \label{RSU_processing_latency_InstrT_Mode_uploading}
    & D_{k,InstrT}^{\mathrm{rsu}} (t)=\frac{G_{i,k}^{\mathrm{InsD}}(t)}{R_{i,k}^{\mathrm{veh\to rsu}}(t)},
    \end{align}

    where $G_{i,k}^{\mathrm{InsD}}(t)$ represents the size of computation instruction allocated to RSU $\mathcal{R}_k$ at time slot $t$.
    
    The uploading time for the InstrT model primarily depends on the coordinate transformation time and $E_{i,k}^{co-tra}(t)$ represents the energy consumption of the RSU executed coordinate transformation. They are given by
    \begin{align}
    \label{RSU_processing_latency_InstrT_Mode_uploading_cotra_and_cotra_energy_consumption}
    & D_{k,InstrT}^{\mathrm{co-tra}} (t)=\frac{\ell_{i,k}^{\mathrm{rsu}}(t) G_{i}^{\mathrm{TasD}}(t) I_{c}^{\mathrm{co-tra}}}{F_{RSU_k}^{\mathrm{cpu}}(t)}, \\
    & E_{i,k}^{\mathrm{co-tra}}(t)= \kappa_{rsu} D_{k,InstrT}^{\mathrm{co-tra}}(t) (F_{RSU_k}^{\mathrm{cpu}}(t))^{3},
    \end{align}

    where $\kappa_{rsu}$ is the energy consumption parameter for RSU computing. $I_{c}^{\mathrm{co-tra}}$ (in cycles/bit) is the computation intensity of the coordinate transformation. $F_{RSU_k}^{\mathrm{cpu}}(t)$ (in cycles/s) denotes the computation resources of the RSU $\mathcal{R}_k$ allocated to the offloaded task at time slot $t$.

    The DT space will choose the one with smaller uploading time between DataT mode and InstrT model. Hence, the uploading time can be defined by
    \begin{equation}
    \label{uploading_to_RSU_processing_latency}
    D_{i,k}^{\mathrm{up}}(t)=\mathrm{min}\{D_{k,DataT}^{\mathrm{rsu}}(t), D_{k,InstrT}^{\mathrm{co-tra}}(t)\}.
    \end{equation}
    
    \item \textit{Computing Time:} After uploading, tasks offloaded to the RSU $\mathcal{R}_k$ will be processed in parallel on the RSU $\mathcal{R}_k$. The RSU processing latency $D_{i,k}^{compt}(t)$ and energy consumption of RSU executing offloading task $E_{i,k}^{\mathrm{compt,rsu}}(t)$ are described as
    \begin{align}
    \label{RSU_processing_computation_latency_and_energy_consumption}
    & D_{i,k}^{\mathrm{compt}}(t)=\frac{\ell_{i,k}^{\mathrm{rsu}}(t) G_{i}^{\mathrm{TasD}}(t) I_{c}^{\mathrm{task}}}{F_{RSU_k}^{\mathrm{cpu}}(t)}, \\
    & E_{i,k}^{\mathrm{compt,rsu}}(t)= \kappa_{rsu} D_{i,k}^{\mathrm{compt}}(t) (F_{RSU_k}^{\mathrm{cpu}}(t))^{3}.
    \end{align}

    To summarize, the total latency of edge processing on the RSU $\mathcal{R}_k$ is given by
    \begin{equation}
    \label{RSU_processing_total_latency}
    D_{i,k}^{\mathrm{rsu}}(t)=D_{i,k}^{\mathrm{up}}(t)+D_{i,k}^{\mathrm{compt}}(t).
    \end{equation}
    
\end{itemize}

\subsection{Queuing Model}\label{subsec35}
In our system, to accurately capture the computational load dynamics across heterogeneous nodes. We establish independent task queues for CVs and RSUs. The queue evolution for each type of node is governed by the discrete-time dynamics based on task arrivals and the available computation resource within time slot duration $\tau$.
\begin{itemize}
    \item \textit{Local Queue} 
    
    For each CV $\mathcal{C}_i$, the local task queue is maintained privately and only accounts for the portion of tasks processed locally. The queue is updated as:   
    \begin{align}
    \label{Local_queue}
    & Q_{i}^{\mathrm{loc}}(t+1)\!=\!\mathrm{max}\{Q_{i}^{\mathrm{loc}}(t)\!-\!\underset{(\mathrm{I})}{\underbrace{C_{CV_i}^{\mathrm{cpu}}\tau}} \!\!+\!\! \underset{(\mathrm{II})}{\underbrace{\lambda _{i}^{\mathrm{loc}}(t)}},0\},
    \end{align}

    where $Q_{i}^{\mathrm{loc}}(t)$ is the backlog of computations in the local task queue of CV $\mathcal{C}_i$ at time slot $t$. The term $(\mathrm{I})$ indicates the amount of computation locally executable within time slot duration $\tau$, and the term $(\mathrm{II})$ represents the newly arrived computations of tasks assigned to the local, therein $\lambda _{i}^{\mathrm{loc}}(t)=\ell_{i}^{\mathrm{loc}}(t) G_{i}^{\mathrm{TasD}}(t) I_{c}^{\mathrm{task}}$.
    
    \item \textit{RSU Queue }

    In contrast, RSUs as shared edge computing nodes and may simultaneously receive offloaded task segments from multiple CV $\mathcal{C}_i$. The computation queues maintained by RSU $\mathcal{R}_k$ must aggregate all arriving task arrivals from different CVs:
    \begin{align}
    \label{RSU_queue}
    Q_{k}^{\mathrm{rsu}}(t+1) &= \mathrm{max}\{Q_{k}^{\mathrm{rsu}}(t)-F_{RSU_k}^{\mathrm{cpu}}(t)\tau \notag \\ 
    & +\underset{(\mathrm{I})}{\underbrace{\sum_{i=1}^{I} \lambda _{i}^{\mathrm{rsu}}(t)}}, 0 \},
    \end{align}

    where $Q_{k}^{\mathrm{rsu}}(t)$ denotes the backlogged computations at the task queue of the RSU $\mathcal{R}_k$ and the term $(\mathrm{I})$ represents the newly arrived computations from all CV $\mathcal{C}_i$ that offload their task to the RSU $\mathcal{R}_k$ at time slot $t$. Here $ \sum_{i=1}^{I} \lambda _{i}^{\mathrm{rsu}}(t) =\ell_{i,k}^{\mathrm{rsu}}(t) G_{i}^{\mathrm{TasD}}(t) [\eta_{i}^{\mathrm{rsu}}(t)I_{c}^{\mathrm{co-tra}}+I_{c}^{\mathrm{task}}]$, therein, $\eta_{i}^{rsu}(t) \in \{0,1\}$ is a mode selector. According to (\ref{uploading_to_RSU_processing_latency}), if InstrT mode has lower total uploading time, $\eta_{i}^{\mathrm{rsu}}(t)$ will be equal to $1$, and otherwise $\eta_{i}^{\mathrm{rsu}}(t)=0$.
    
\end{itemize}

After modeling the evolution of task queues at CVs and RSUs, we proceed to estimate the corresponding queueing delay encountered by tasks assigned to each node. In order to capture the average waiting time before execution, we apply Little’s Law, a widely adopted principle in queuing theory, which states that the expected delay in a stable system is proportional to the average number of tasks in the queue divided by the average arrival rate \cite{9449944}.

Specifically, we define the queueing delay at a given node $ n \in \{\mathrm{loc,rsu}\} $ for a task originated by CV $\mathcal{C}_i$ as the ratio between the time-averaged queue length and the time-averaged arrival computation load. 

Let $Q_{i}^{\mathrm{n}}(t)$ and $\lambda_{i}^{\mathrm{n}}(t)$ denote the queue length and task arrival load at time slot $t$, the queueing delay is expressed as:
\begin{equation}
\label{queue_latency}
D_{i}^{\mathrm{queue,n}}(t) = \frac{\widetilde{Q_{i}^{\mathrm{n}}}(t)}{\widetilde{\lambda_{i}^{\mathrm{n}}}(t)},
\end{equation}
where $\widetilde{Q_{i}^{\mathrm{n}}}(t)=\frac{1}{m}\sum_{l=t-m+1}^{t}Q_{i}^{\mathrm{n}}(l)$ and $\widetilde{\lambda_{i}^{\mathrm{n}}}(t)=\frac{1}{m}\sum_{l=t-m+1}^{t}\lambda_{i}^{\mathrm{n}}(l)$ are the moving average queue length and arrival rate computed over a sliding time window of size $m$, respectively.

Consequently, $T_{i}^{\mathrm{task}}(t)$ (i.e., the maximum of local partial latency and RSU partial latency) is taken as the overall latency of all tasks executed in parallel. $E_i^{\mathrm{task}}(t)$ denotes the total energy consumed by all nodes (for both CVs and RSUs) involved in executing the task initiated by $\mathcal{C}_i$ at time slot $t$. They can be defined as follows:
\begin{align}
\label{overall_latency_and_total_energy_consumption}
T_{i}^{\mathrm{task}}(t)&= \mathrm{max} \{
T_{i}^{\mathrm{loc}}(t),
T_{i,k}^{\mathrm{rsu}}(t)\}, \\
E_{i}^{\mathrm{task}}(t) & =E_{i}^{\mathrm{loc}}(t) \notag \\ 
& + \sum_{k=1}^{K}[E_{i,k}^{\mathrm{up,rsu}}(t)+E_{i,k}^{\mathrm{compt,rsu}}(t)], 
\end{align}
where $T_{i}^{\mathrm{loc}}(t)=D_{i}^{\mathrm{loc}}(t) + D_{i}^{\mathrm{queue,loc}}(t)$ is the local part total latency. $T_{i,k}^{\mathrm{rsu}}(t)=D_{i,k}^{\mathrm{rsu}}(t) + D_{i}^{\mathrm{queue,rsu}}(t),  \mathcal{R}_k \in  \mathbb{R} $ denotes the RSU part total latency.

Combining (\ref{local_processing_latency_and_energy_consumption}), (\ref{RSU_processing_total_latency}), and (\ref{queue_latency}), the overall latency is updated by
\begin{align}
\label{overall_latency_updated}
& T_{i}^{\mathrm{task}}(t) = \mathrm{max} \{T_{i}^{\mathrm{loc}}(t), T_{i,k}^{\mathrm{rsu}}(t)\}, \notag \\
& T_{i}^{\mathrm{loc}}(t) = \frac{\ell_{i}^{\mathrm{loc}}(t) G_{i}^{\mathrm{TasD}}(t) I_{c}^{\mathrm{task}}}{C_{CV_i}^{\mathrm{cpu}}} + \frac{\widetilde{Q_{i}^{loc}}(t)}{\widetilde{\lambda_{i}^{loc}}(t)}, \notag \\
& T_{i,k}^{\mathrm{rsu}}(t)= \mathrm{min}\{\frac{\ell_{i,k}^{\mathrm{rsu}}(t) G_{i}^{\mathrm{TasD}}(t)}{R_{i,k}^{\mathrm{veh\to rsu}}(t)}, \frac{\ell_{i,k}^{\mathrm{rsu}}(t) G_{i}^{\mathrm{TasD}}(t) I_{c}^{\mathrm{co-tra}}}{F_{RSU_k}^{\mathrm{cpu}}(t)}\} \notag \\ 
& + \frac{\ell_{i,k}^{\mathrm{rsu}}(t) G_{i}^{\mathrm{TasD}}(t) I_{c}^{\mathrm{task}}}{F_{RSU_k}^{\mathrm{cpu}}(t)} + \frac{\widetilde{Q_{i}^{\mathrm{rsu}}}(t)}{\widetilde{\lambda_{i}^{\mathrm{rsu}}}(t)}(\ 1 \le k \le K).
\end{align}

The energy consumption for uploading tasks depends on the selected transmission mode. To ensure consistency, we refine the mode selector $\eta_{i}^{\mathrm{n'}}(t)$ mentioned in Sec. \ref{subsec35} and combine (\ref{uploading_to_RSU_processing_latency}), defined as
\begin{align}
\label{Mode_Selector}
\eta_{i}^{\mathrm{rsu}}(t) =\left\{
\begin{aligned}
& 1, D_{i,DataT}^{\mathrm{rsu}}(t) < D_{i,InstrT}^{\mathrm{co-tra}}(t) \\
& 0, otherwise
\end{aligned} \right. .
\end{align}
Hence, the total energy consumption can be updated by 
\begin{align}
\label{total_energy_consumption_updated}
E_{i}^{\mathrm{task}}(t) & =E_{i}^{\mathrm{loc}}(t) \notag \\ 
& + \sum_{k}[(1-\eta_{i}^{\mathrm{rsu}}(t))E_{i,k}^{\mathrm{DataT-up}}(t) \notag \\
& +  \eta_{i}^{\mathrm{rsu}}(t)E_{i,k}^{\mathrm{co-tra}}(t) + E_{i,k}^{\mathrm{compt,rsu}}(t)].
\end{align}

\section{Problem Formulation}\label{sec4}
To achieve a practical and interpretable system objective, we formulate the optimization problem in terms of total cost, which jointly considers the task latency and the system-wide energy consumption incurred to support each task. Specifically, for each client vehicle $\mathcal{C}_i$, we define its service-related cost as:
\begin{equation}
\label{individual_cost_Function}
\complement_{i}(t) = \alpha T_{i}^{\mathrm{task}}(t) + (1-\alpha)E_{i}^{\mathrm{task}}(t),
\end{equation}
where $\alpha \in [0,1]$ is a configurable weight to flexibly balance latency and energy consumption. 

Based on this, the system-level cost is defined as the sum over all CVs as follows:
\begin{equation}
\label{System_Cost_Function}
\complement(t) = \sum_{i \in \mathbb{C}} \complement_{i}(t).
\end{equation}

The system-wide objective is then to minimize the cumulative cost over all CVs, which can be formulated as
\begin{align}
\label{P1_problem}
& {\mathcal{P}_1:} \quad
\min_{\{\textbf{b}(t), \ell(t), F(t) \}} 
\lim_{T \to +\infty} \frac{1}{T} \sum_{t \in \mathcal{T}} \mathbb{E}[\complement(t)], \\
\text{s.t.} \quad
& \text{C1\ :} \lim_{T \to +\infty} \frac{1}{T} \sum_{t \in \mathcal{T}} \mathbb{E}[Q_i^{\mathrm{loc}}(t)] < \infty,\quad \forall \mathcal{C}_i \in \mathbb{C}, \notag\\
& \text{C2\ :} \lim_{T \to +\infty} \frac{1}{T} \sum_{t \in \mathcal{T}} \mathbb{E}[Q_k^{\mathrm{rsu}}(t)] < \infty,\quad \forall \mathcal{R}_k \in \mathbb{R}, \notag\\
& \text{C3\ :} \lim_{T \to \infty} \frac{1}{T} \sum_{t \in \mathcal{T}} \mathbb{E}[E^{\mathrm{total}}(t)] \le E^{\mathrm{max}}, \notag\\
& \quad \ \ \ E^{\mathrm{total}}(t) = \sum_{\mathcal{C}_i \in \mathbb{C}} E_i^{\mathrm{task}}(t), \notag\\
& \text{C4\ :} \ \ 0 \le b_{i,k}^{\mathrm{veh\to rsu}}(t) \le 1,\quad \forall \mathcal{R}_k \in \mathbb{R}, \notag\\
& \quad \ \ \sum_{\mathcal{R}_k \in \mathbb{R}} b_{i,k}^{\mathrm{veh\to rsu}}(t) \le 1,\quad \forall \mathcal{C}_i \in \mathbb{C}, \forall t \in \mathcal{T}, \notag\\
& \text{C5\ :} \ \ 0 \le \ell_i^{\mathrm{loc}}(t),\; \ell_{i,k}^{\mathrm{rsu}}(t) \le 1,\quad \forall \mathcal{R}_k, \notag\\
& \quad \ \ \ \ell_i^{\mathrm{loc}}(t) + \! \sum_{\mathcal{R}_k \in \mathbb{R}} \ell_{i,k}^{\mathrm{rsu}}(t) = 1,\quad \forall \mathcal{C}_i \in \mathbb{C}, \forall t \in \mathcal{T}, \notag\\
& \text{C6\ :} \ \ 0 \le F_{RSU_k}^{\mathrm{cpu}}(t) \le F_k^{\mathrm{rsu,max}} - C_k^{\mathrm{twin}}(t), \notag \\
& \quad \ \ \ \forall \mathcal{R}_k \in \mathbb{R}, \forall t\in \mathcal{T}, \notag\\
& \text{C7\ :} \ \ \eta_i(t) \in \{0,1\},\quad \forall \mathcal{C}_i \in \mathbb{C}, \forall t\in \mathcal{T}, \notag\\
& \quad \ \ \ \text{(0: DataT mode, 1: InstrT mode)}, \notag\\
& \text{C8\ :} \ \ T_i(t) \le T_i^{\max},\quad \forall \mathcal{C}_i \in \mathbb{C}, \forall t\in \mathcal{T}. \notag
\end{align}
where the optimization variables are (i) $\textbf{b}(t)\!\!=\{b_{i,k}(t) \}_{\mathcal{C}_i \in \mathbb{C},\mathcal{R}_k \in \mathbb{R}}$ is the bandwidth allocation ratios between CV $\mathcal{C}_i$ and RSU $\mathcal{R}_k$, at time slot $t$; (ii) $ \ell(t)=\{\ell_i^{\mathrm{loc}}(t), \ell_{i,k}^{\mathrm{rsu}}(t)\}_{\mathcal{C}_i \in \mathbb{C},\mathcal{R}_k \in \mathbb{R}}$ is the task assignment ratio, indicating how the task of $\mathcal{C}_i$ is distributed among local processing and RSU nodes; (iii) $F(t)=\{F_{RSU_k}^{\mathrm{cpu}}(t)\}_{\mathcal{R}_k \in \mathbb{R}}$ is the RSU computing resource allocation for subtasks at time $t$.

In $(\mathcal{P}_1)$, constraints $\text{C1}$ and $\text{C2}$ ensure the long-term stability of task queues at the CVs and RSUs, respectively. Constraint $\text{C3}$ is the long-term average total system energy consumption to be within a predefined system-wide energy budget $E^{\mathrm{max}}$. Therein, $E_i^{\mathrm{task}}(t)$ represents the energy consumption caused by a task initiated by CV $\mathcal{C}_i$ in the entire system, and $E^{\mathrm{total}}(t)$ denotes the total system energy consumption caused by all tasks initiated by CVs in time slot $t$. Constraint $\text{C4}$ \cite{10638833} ensures that the bandwidth allocation ratios confined within valid bounds and that the total allocated bandwidth for each CV $\mathcal{C}_i$ at any given time slot does not exceed the normalized unit bandwidth. Constraint $\text{C5}$ guarantees that the task assignment proportions across all available computing nodes sum to one, with each individual ratio constrained to the interval $[0,1]$. Constraint $\text{C6}$ limits the computing resources allocated by RSUs to ensure that the assigned resources do not exceed the remaining available capacity after reserving a portion $C_k^{\mathrm{twin}}(t)$ for constructing the DT space. Constraint $\text{C7}$ represents the binary mode selector $\eta_i(t) \in \{0,1\}$, where 0 denotes DataT mode and 1 denotes InstrT mode for CV $\mathcal{C}_i$. Constraint $\text{C8}$ ensures that the task execution delay perceived by the CV does not exceed its maximum tolerable latency $T_i^{\max}$.

\begin{remark}
The joint optimization problem $(\mathcal{P}_1)$ involves long-term constraints on queue stability $\text{C1}$ and $\text{C2}$, which inherently couple the task offloading and resource allocation decisions across different time slots. This temporal coupling arises from the dynamic evolution of task queues over time, making the decision space sequentially dependent. Additionally, the coexistence of continuous variables (e.g., task and bandwidth allocation ratios in constraints $\text{C4}$ and $\text{C5}$, and discrete decision variables (e.g., computation mode selector in \text{C7} results in an MINLP structure. Such problems are generally NP-hard and difficult to solve efficiently in real-time, especially under dynamic vehicle mobility and limited prior knowledge of future states.
\end{remark}

\section{Lyapunov-Based Dynamic Long-Term Problem Decoupling} \label{sec5}
To tackle this challenge mentioned above, we employ a Lyapunov optimization approach to reformulate the original long-term problem into a series of tractable per-slot decisions. This enables real-time decision-making without requiring complete knowledge of future system dynamics, while still providing long-term guarantees on the stability and feasibility constraints.

In Lyapunov optimization, physical task queues ($\text{C1}$ and $\text{C2}$) are directly incorporated into the Lyapunov function and drift expressions to ensure long-term queue stability. In contrast, for system-level time-average constraints such as energy consumption $\text{C3}$, which is not related to explicit queueing structures, virtual queues must be introduced to transform these average constraints into queue stability conditions. To handle the long-term system-wide energy constraint $\text{C3}$, we introduce a virtual energy queue $V_{i}(t)$ that tracks the cumulative deviation of actual energy usage from the permissible average. Although the time slot index set is defined as $\mathcal{T} = \{1, \ldots, T\}$, for convenience of Lyapunov analysis, we adopt $t = 0$ as the initial slot and set the virtual queue state $V_{i}(0) = 0$.

The virtual queue evolves according to
\begin{equation}
\label{virtual_queues}
V_{i}(t+1)=\mathrm{max}\{V_{i}(t) - E^{\mathrm{max}} + E^{\mathrm{total}}(t),0\}, 
\end{equation}
therein, stabilizing $V_{i}(t)$ guarantees that the time-average energy consumption constraint is satisfied in the long run. 

To examine the long-term behavior of $V_{i}(t)$, we first relax the max operator for tractability and analyze the inequality:
\begin{equation}
V_{i}(t+1) - V_{i}(t) \ge E^{\mathrm{total}}(t) - E^{\max}
\label{virtual_queue_diff}.
\end{equation}

Summing both sides of \eqref{virtual_queue_diff} from $t = 0$ to $T-1$, we obtain:
\begin{equation}
V_{i}(T) - V_{i}(0) \ge \sum_{t=0}^{T-1} \left( E^{\mathrm{total}}(t) - E^{\max} \right).
\end{equation}

Dividing both sides by $T$ and rearranging gives:
\begin{equation}
\frac{1}{T} \sum_{t=0}^{T-1} E^{\mathrm{total}}(t) \le E^{\max} + \frac{V_{i}(T) - V_{i}(0)}{T}.
\end{equation}

Taking expectations and noting $V_{i}(0) = 0$, we derive:
\begin{equation}
\frac{1}{T} \sum_{t=0}^{T-1} \mathbb{E}[E^{\mathrm{total}}(t)] \le E^{\max} + \frac{\mathbb{E}[V_{i}(T)]}{T}.
\label{energy_expectation_bound}
\end{equation}

Hence, if the virtual queue $V_{i}(t)$ is mean-rate stable, i.e.,
\begin{equation}
\lim_{T \to \infty} \frac{\mathbb{E}[V_{i}(T)]}{T} = 0.
\end{equation}

Then, the original energy consumption constraint $\text{C4}$ is satisfied \cite{10638833}:
\begin{equation}
\lim_{T \to \infty} \frac{1}{T} \sum_{t=0}^{T-1} \mathbb{E}[E^{\mathrm{total}}(t)] \le E^{\max}.
\end{equation}

Therefore, stabilizing the virtual queue $V_{i}(t)$ ensures the satisfaction of the long-term average energy consumption constraint $\text{C3}$.

We define $\mathbf{Z} (t)$ as a compact representation of the backlog state at time slot $t$, which includes the task queues of the CVs and the RSUs, as well as the virtual queues $V_{i}(t)$: 
\begin{equation}
\mathbf{Z} (t) = \{ \{ Q_{i}^{\mathrm{loc}}(t)\}_{\mathcal{C}_i \in \mathbb{C}},\{Q_{k}^{\mathrm{rsu}}(t)\}_{\mathcal{R}_k \in \mathbb{R}}, V_{i}(t) \}. 
\end{equation}

We define the Lyapunov function as the scalar quadratic measure of queue backlogs:
\begin{align}
\label{Lyapunov_function}
L(\mathbf{Z}(t)) 
= \frac{1}{2} \Bigg[
& \sum_{i \in \mathbb{C}} \left( Q_{i}^{\mathrm{loc}}(t) \right)^2 
+ \sum_{k \in \mathbb{R}} \left( Q_{k}^{\mathrm{rsu}}(t) \right)^2 \notag \\
& + \sum_{i \in \mathbb{C}}V_{i}(t)^2
\Bigg].
\end{align}

The Lyapunov drift, characterizing the changes in queue states over successive time slots, is derived by applying the Lyapunov function in (\ref{Lyapunov_function}) across two consecutive slots:
\begin{equation}
\label{Lyapunov_drift_function}
\Delta(\mathbf{Z}(t)) = \mathbb{E} \left[ L(\mathbf{Z}(t+1)) - L(\mathbf{Z}(t)) \mid \mathbf{Z}(t) \right].
\end{equation}

Leveraging the Lyapunov drift-plus-penalty framework, we formulate a drift-penalty function that jointly incorporates the Lyapunov drift and the objective function of problem $(\mathcal{P}_1)$, aiming to minimize the following upper bound of the drift-plus-penalty expression at each time slot as follows:
\begin{equation}
\label{Drift_plus_penalty}
\Lambda (\mathbf{Z}(t))=\Delta(\mathbf{Z}(t)) + V \cdot \mathbb{E}[\complement(t) \mid \mathbf{Z}(t)],
\end{equation}
where $V > 0 $ is the weight parameter that balances the trade-off between minimizing control cost and ensuring queue stability.

In the following, we obtain an upper bound of $\Lambda (\mathbf{Z}(t))$ on (\ref{Drift_plus_penalty}).

\textit{Lemma 1:} For any feasible set of $\{ \textbf{b}(t), \ell(t), F(t) \}_{t\in \mathcal{T}}$, which satisfies constraints $\text{C1} \sim \text{C8}$, the Lyapunov drift-plus-penalty expression $\Lambda (\mathbf{Z}(t))$ can be upper bounded as follows:
\begin{align}
& \Lambda (\mathbf{Z}(t)) = \Delta(\mathbf{Z}(t)) + V \cdot \mathbb{E}[\complement(t) \mid \mathbf{Z}(t)] \notag \\
& \le \mathbb{E} \bigg[ \sum_{\mathcal{C}_i\in \mathbb{C}} Q_{i}^{\mathrm{loc}}(t)({\lambda _{i}^{\mathrm{loc}}(t)} - {C_{CV_i}^{\mathrm{cpu}}\tau}) \notag \\
& + \sum_{\mathcal{R}_k\in \mathbb{R}} Q_{k}^{\mathrm{rsu}}(t) (\sum_{\mathcal{C}_i\in \mathbb{C}} \lambda _{i}^{\mathrm{rsu}}(t) - F_{RSU_k}^{\mathrm{cpu}}(t)\tau) \notag \\
& + \sum_{\mathcal{C}_i\in \mathbb{C}} V_{i}(t) (E^{\mathrm{total}}(t) - E^{\mathrm{max}}) \notag \\
& + V \cdot \complement(t)
\ \bigg| \ \mathbf{Z}(t) \bigg] + B,
\end{align}
where $B$ is a constant upper bound on the second-order moments of queue variations, determined by the per-time slot decisions $\{ \textbf{b}(t), \ell(t), F(t) \}_{t\in \mathcal{T}}$. 

The detailed proof is attached to Appendix A, available online.

Based on the upper bound in Lemma 1, the drift-plus-penalty expression $ \Lambda (\mathbf{Z}(t)) $ can be minimized by solving a deterministic optimization problem at each time slot.

Since the constant term $B$ does not influence the decision-making process, it can be safely omitted from the objective function. Accordingly, we reformulate problem ($\mathcal{P}_1$) as a per-slot problem ($\mathcal{P}_2$), which can be solved online with current observations, ensuring long-term constraint satisfaction and queue stability.
\begin{align}
\label{P2_problem}
\mathcal{P}_2: \quad 
& \min_{\{ \textbf{b}(t), \ell(t), F(t) \}} 
V \cdot \complement(t) \\
& + \sum_{\mathcal{C}_i\in \mathbb{C}} Q_{i}^{\mathrm{loc}}(t)({\lambda _{i}^{\mathrm{loc}}(t)} - {C_{CV_i}^{\mathrm{cpu}}\tau}) \notag \\
& + \sum_{\mathcal{R}_k\in \mathbb{R}} Q_{k}^{\mathrm{rsu}}(t) (\sum_{\mathcal{C}_i\in \mathbb{C}} \lambda _{i}^{\mathrm{rsu}}(t) - F_{RSU_k}^{\mathrm{cpu}}(t)\tau) \notag \\
& + \sum_{\mathcal{C}_i\in \mathbb{C}} V_{i}(t) (E^{\mathrm{total}}(t) - E^{\mathrm{max}}) \notag \\
\text{s.t.} \quad 
& \text{Constraints} \quad \text{C4} \sim \text{C8} \quad \text{in} \quad (\mathcal{P}_1). \notag
\end{align}

\section{Lyapunov-Based DRL for Resource Allocation} \label{sec6}
In this section, we cast the per-slot problem into a Markov decision process (MDP). We then develop the CTDE actor–critic solution, detailing its MAPPO-based architecture and training procedure, followed by a computational complexity analysis that quantifies the cost of trajectory collection and actor–critic updates.

\subsection{Markov Decision Process}
To solve the derived per-slot optimization problem ($\mathcal{P}_{2}$) in an online manner under dynamic vehicular environments and uncertain network conditions, we further formulate it as an MDP. This MDP formulation enables the application of DRL to learn optimal offloading and resource allocation strategies through interaction with the environment. While recent studies have incorporated diffusion models into deep deterministic policy gradient frameworks to enhance exploration efficiency in high-dimensional continuous spaces \cite{11223083}, our approach differs by embedding Lyapunov drift–plus–penalty theory into the reward design, explicitly guaranteeing queue stability and long-term energy efficiency in dynamic vehicular environments. Specifically, the MDP is defined by the following components: state space, action space, and reward function.

\textit{1) State Space:} In our system, each CV $\mathcal{C}_i$ is modeled as an individual agent. At each time slot $t$, we adopt a DT-enhanced state description incorporating vehicle and RSU dynamics, as well as additional queueing states introduced in our Lyapunov-based model. The local state $s_i(t)$ is given by
\begin{equation}
\label{state space}
\!\!\!s_i(t)=\{\mho_{info}^{\mathrm{CV}}, \mho_{info}^{\mathrm{RSU}}, Q_{i}^{\mathrm{loc}}(t), Q_{k}^{\mathrm{rsu}}(t), V_{i}(t),G_{rsu}(t)\},
\end{equation}
therein, the DT of CV $\mathcal{C}_i$ can be defined as $\mho_{info}^{\mathrm{CV}}=\{ G_{i}^{\mathrm{TasD}}(t), G_{i}^{\mathrm{InsD}}(t), C_{CV_i}^{\mathrm{cpu}}, L_{i}^{\mathrm{cv}}(t), v_{i}^{\mathrm{cv}}, T_{i}^{\mathrm{max}}\}$. The DT of RSU $\mathcal{R}_k$ can be defined as $\mho_{info}^{\mathrm{RSU}}=\{ L_{k}^{\mathrm{rsu}}, F_k^{\mathrm{rsu,max}}\}$.

Then, the joint state of all agents is defined as follows:
\begin{equation}
S(t)=\{ s_1(t),\ldots,s_i(t),\ldots,s_I(t)\}.
\end{equation}

\textit{2) Action Space:} The agent $i$, corresponding to the CV $\mathcal{C}_i$, is responsible for making task offloading and bandwidth allocation decisions. Specifically, after a task is generated at the beginning of time slot $t$, the DT space provides real-time environment and resource information to assist the decision-making process. Based on this, the agent determines: (i) the partition ratios of the task to be processed locally, and offloaded to RSU nodes; and (ii) the bandwidth allocation ratios for each offloading link. Therefore, the action space of agent $i$ at time slot $t$ is defined as:
\begin{equation}
a_i(t)=\{ \textbf{b}(t), \ell(t), F(t)\},
\end{equation}
where $\textbf{b}(t)$ satisfies constraint \text{C4} and $\ell(t)$ satisfies constraint \text{C5}. 

It is important to note that the transmission mode selection variable $\eta_i(t)$, is not directly decided by the agent. Instead, it is automatically determined by the DT space. $\eta_i(t)$ satisfies constraint \text{C7}.

Then, the joint action of all agents is defined as below:
\begin{equation}
a(t)=\{ a_1(t),\ldots,a_i(t),\ldots,a_I(t)\}.
\end{equation}

\textit{3) Reward Function:} After executing action $a_i(t)$ at state $s_i(t)$, agent $i$ observes the transition of the environment to a new state $s_i(t + 1)$, and receives a reward that reflects the quality of the decision. In our system, the reward function is derived from the drift-plus-penalty formulation obtained via Lyapunov optimization. Specifically, it is designed as the negative of the transformed per-slot objective function in Problem ($\mathcal{P}_2$), which jointly captures queue stability, task completion delay, and energy consumption \cite{10638833}. This design ensures that maximizing the expected cumulative reward is equivalent to minimizing the long-term system cost while maintaining queue stability.

The reward function is therefore defined as:
\begin{align}
\label{reward_function}
R(t)
& =- \Bigg\{ 
V \cdot \complement(t) + \sum_{\mathcal{C}_i\in \mathbb{C}} Q_{i}^{\mathrm{loc}}(t)({\lambda _{i}^{\mathrm{loc}}(t)} - {C_{CV_i}^{\mathrm{cpu}}\tau}) \notag \\
& + \sum_{\mathcal{R}_k\in \mathbb{R}} Q_{k}^{\mathrm{rsu}}(t) (\sum_{\mathcal{C}_i\in \mathbb{C}} \lambda _{i}^{\mathrm{rsu}}(t) - F_{RSU_k}^{\mathrm{cpu}}(t)\tau) \notag \\
& + \sum_{\mathcal{C}_i\in \mathbb{C}} V_{i}(t) (E^{\mathrm{total}}(t) - E^{\mathrm{max}})
\Bigg\}.
\end{align}

\subsection{DT-Enhanced DRL Algorithm}
To solve the Lyapunov-optimized per-slot decision problem derived in ($\mathcal{P}_2$), this problem is modeled as a cooperative multi-agent reinforcement learning (MARL) problem with shared resource constraints. Although all agents aim to cooperatively minimize a system-wide Lyapunov drift-plus-penalty objective, the agents implicitly compete for limited RSU computation and bandwidth resources. Therefore, this scenario constitutes a soft-coupled cooperative MARL problem, where agents are cooperative in policy optimization but interdependent due to constrained and shared resource allocation.

\textit{1) The Motivation of Adopting MAPPO \cite{9895362,10620870}:} Although Lyapunov optimization effectively decomposes the long-term coupled optimization problem into tractable per-slot subproblems, solving the resulting mixed-integer nonlinear programming problem in real time remains challenging due to the high-dimensional decision space and rapidly changing vehicular environments. Traditional optimization and heuristic approaches are often insufficient to capture the complex dynamics of joint resource allocation and task offloading under mobility, heterogeneity of tasks, and stringent latency constraints. To overcome these challenges, we adopt the MAPPO framework, which naturally fits the cooperative yet resource-competitive characteristics of vehicular networks. Within the CTDE paradigm, MAPPO utilizes a centralized critic to leverage global information during training, while allowing decentralized actors to execute policies independently in real-time. Moreover, by embedding the Lyapunov drift-plus-penalty formulation into the reward function, the algorithm explicitly incorporates queue stability, latency, and energy trade-offs into the learning process. This integration ensures not only adaptability to dynamic network conditions but also long-term system stability. 

\begin{figure*}[h]
\centering
\includegraphics[width=7in]{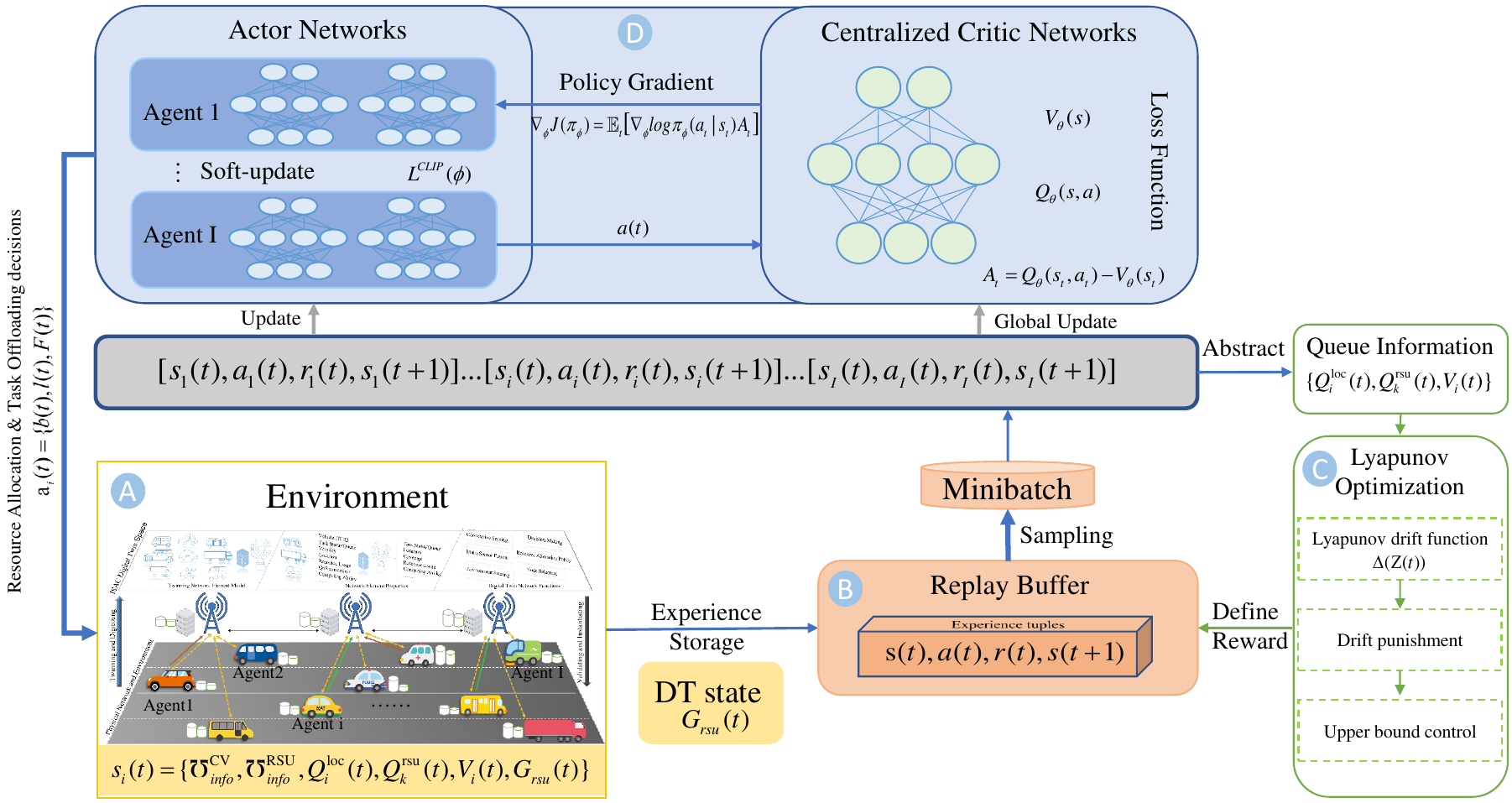}
\caption{Framework of the proposed Ly-DTMPPO algorithm. Part A describes the DT-assisted vehicular environment where agents interact with the system. Part B stores and samples experiences for policy updates. Part C employs Lyapunov optimization to stabilize queue dynamics and balance delay–energy trade-offs. Part D adopts a CTDE structure with centralized critics and decentralized actors for cooperative policy learning.} 
\label{fig_3}
\end{figure*}

\textit{2) Network Architecture of Ly-DTMPPO:} As shown in Fig. \ref{fig_3}, the proposed Ly-DTMPPO framework adopts an actor–critic architecture under the CTDE paradigm. Each agent is equipped with an actor network for decentralized decision making, while a centralized critic network is employed during training to guide policy improvement using global information.

\begin{itemize}
    \item \textbf{\textit{Actor Network:}}
    For each agent $i \in \mathcal{I} $, the actor network takes its local observation $s_{i}(t)$, which includes the DT-enhanced vehicular state (e.g., task size, CPU capability, position, and velocity) and queue-related information and the RSU-level queue aggregation $G_{\mathrm{rsu}}(t) = [\sum_{k=1}^{K} Q^{\mathrm{rsu}}_k(t), \tfrac{1}{K}\sum_{k=1}^{K} Q^{\mathrm{rsu}}_k(t), \max_k Q^{\mathrm{rsu}}_k(t)]$. When DT information is unavailable, this component is replaced with a zero vector of the same dimension to keep the state representation consistent. The actor consists of two fully connected hidden layers with ReLU activations, followed by an output layer that parameterizes a Gaussian policy. Specifically, the actor outputs the mean and variance of the action distribution, from which the offloading and resource allocation decision $a_{i}(t)$ is sampled. To stabilize the policy optimization, the PPO objective is used, where the clipped surrogate loss $\mathit{L}^{CLIP}(\phi)$ ensures conservative policy updates. During execution, each agent selects its action in a distributed manner based solely on its local state.
    \item \textbf{\textit{Critic Network:}}
    The critic network is centralized and shared during training. It takes as input the global state $s(t)$, which aggregates the observations of all agents and system-level queue information, and the RSU queue aggregation $G_{\mathrm{rsu}}(t)$. It outputs the state-value estimate $\mathit{V}_{\theta}(s(t))$. The critic consists of two fully connected hidden layers with ReLU activations and one linear output layer. By minimizing the mean squared error between predicted and empirical returns, the critic provides stable value function approximation. The advantage function is then derived as
    \begin{equation}
    \label{advantage_function}
    {A_t} = Q_{\theta}(s_t,a_t) - V_{\theta}(s_t),
    \end{equation}
    which is used to guide the actor updates via policy gradient.
\end{itemize}

\textit{3) Proposed Ly-DTMPPO algorithm description:} We concretely introduce the proposed Ly-DTMPPO for each CV $\mathcal{C}_i$ in Algorithm \ref{alg:Ly-DTMPPO}. The overall training and execution procedure is summarized as follows.

\begin{itemize}
    \item \textbf{\textit{Trajectory Collection:}} At each decision slot $t$, every agent $i \in \mathcal{I}$ observes its local state $s_{i}(t)$ and generates an action $a_{i}(t)$ according to its stochastic policy $\pi _\phi (a_i|s_i)$. The joint action $a(t)=\{a_1(t),\ldots,a_I(t)\}$ is executed in the environment, which then returns the next state $s(t+1)$ and reward vector $r(t)=\{r_1(t),\ldots,r_I(t)\}$. The reward is defined by the Lyapunov drift-plus-penalty formulation to simultaneously capture delay, energy, and queue stability constraints. The experience tuple $(s(t),a(t),r(t),s(t+1))$ is stored in the replay buffer.
    
    \item \textbf{\textit{Advantage Estimation:}} For each agent, the critic network approximates the value function $V_\theta (s_t)$. To compute the temporal-difference error and stabilize training, we adopt generalized advantage estimation (GAE). The advantage for step $t$ is calculated as
    \begin{equation}
        \label{advantage_function_detail}
        A_t = \delta _t +(\gamma \lambda )\delta _{t+1}+\cdots+(\gamma \lambda )^{T-t+1}\delta _{T-1},
    \end{equation}
    where $\delta _t=r_t+\gamma V_{\theta}(s_{t+1})-V_{\theta}(s_t)$, and $\gamma$ and $\lambda$ denote the discount and trace-decay factors, respectively.
    \item \textbf{\textit{Actor Update:}} The actor parameters $\phi$ are optimized using the PPO clipped surrogate objective to prevent excessively large policy updates:
    \begin{equation}
    \label{PPO_clipp}
    L^{CLIP}\!(\phi)\!\! =\!\! \hat {\mathbb{E}_t}[\min(r_t(\phi){A}_t, \text{clip}(r_t(\phi),1-\epsilon,1+\epsilon){A}_t)],
    \end{equation}  
    where $r_t(\phi)= \frac{\pi _\phi (a_t|s_t)}{\pi _{\phi _{old}} (a_t|s_t)} $ is the probability ratio between the new and old policies, and $\epsilon$ is the clipping parameter. The gradient of the actor objective is then computed as
    \begin{equation}
    \label{gradient_computed}
    \nabla_\phi J(\pi_\phi) \approx  {\mathbb{E}_t}[\nabla_\phi log \pi _\phi (a_t|s_t){A}_t].
    \end{equation}  

    \item \textbf{\textit{Critic Update:}} The critic network parameters $\theta$ are updated by minimizing the mean-squared error (MSE) between the predicted value and the empirical return:
    \begin{equation}
    \label{MSE}
    L_V(\theta) =  {\mathbb{E}_t}[(V_\theta(s_t)-R_t)^2],
    \end{equation}  
    where $R_t=A_t+V_\theta(s_t)$ denotes the estimated return at time $t$. The critic thus provides a stable baseline for advantage computation.

    \item \textbf{\textit{Training Loop:}} The actor and critic are updated for multiple epochs using minibatches of trajectories sampled from the replay buffer. To prevent gradient explosion, gradient clipping is applied. An entropy bonus is also added to the actor loss to encourage sufficient exploration in the action space.
    \item \textbf{\textit{Execution Phase:}} After centralized training, each vehicle executes the learned policy independently based on its own local observation $s_i(t)$. This decentralized execution ensures scalability in real vehicular networks, while the Lyapunov-shaped reward guarantees queue stability and energy–latency trade-offs over the long run.
\end{itemize}

\textit{4) Complexity Analysis:} The computational complexity of the proposed Ly-DTMPPO algorithm mainly arises from four components: trajectory collection, advantage estimation, actor–critic updates, and centralized training \cite{11023535,11126101}.

In the \textbf{trajectory collection} phase, each of the $\mathcal{I}$ agents executes its policy network once per time slot, resulting in a complexity of $\mathcal{O}(\mathcal{I} \times d_\pi)$, where $d_\pi$ denotes the number of actor parameters. \textbf{Advantage estimation} introduces a complexity of $\mathcal{O}(\mathcal{I} \times T)$ over $T$ decision steps. For \textbf{actor–critic updates}, mini-batch training with batch size $Bs$ and $E$ epochs yields a total complexity of $\mathcal{O}(E \times Bs \times (d_\pi + d_V))$, where $d_V$ is the number of critic parameters. Under the CTDE paradigm, the communication overhead during centralized training is $\mathcal{O}(I \times d_\pi)$ per synchronization, while decentralized execution incurs only $\mathcal{O}(d_\pi)$ inference cost per agent. Overall, the per-iteration complexity of Ly-DTMPPO can be expressed as
\begin{equation}
    \mathcal{O}(\mathcal{I} \times (T + d_\pi) + E \times Bs \times (d_\pi + d_V)),
\end{equation}
which grows linearly with the number of agents and decision steps. Benefiting from decentralized execution and parallel inference, Ly-DTMPPO achieves scalable and real-time applicability in large-scale vehicular networks.
\begin{algorithm}[t!]
    \caption{Proposed Ly-DTMPPO algorithm for agent $i$ in DT-enabled V2X networks}
    \label{alg:Ly-DTMPPO}
    \renewcommand{\algorithmicrequire}{\textbf{Input:}}
    \renewcommand{\algorithmicensure}{\textbf{Output:}}
    \begin{algorithmic}[1]
    
    \REQUIRE State space $s_i(t)$ defined in (\ref{state space}); actor $\pi_{\varphi_i}$; critic $V_\theta$; buffer $\mathcal{D}$
    \ENSURE Trained actor $\pi_{\varphi_i}$ and critic $V_\theta$
    
    \FOR{each episode}
        \STATE Initialize environment $\mathcal{E}$, obtain initial state $s_i(0)$, clear buffer $\mathcal{D}$
        \FOR{$t=1,2,\ldots,T$}
            \STATE Observe current state $s_i(t)$ as defined in (\ref{state space});
            \STATE Each agent $i$ selects action $a_i(t) \sim \pi_{\varphi_i}(a_i|s_i(t))$;
            \STATE Execute joint action $a(t)=\{a_1(t),\ldots,a_I(t)\}$ in $\mathcal{E}$;
            \STATE Observe next state $s_i(t+1)$ and reward $r_i(t)$ defined by (\ref{reward_function});
            \STATE Store $(s_i(t), a_i(t), r_i(t), s_i(t+1))$ into $\mathcal{D}$;
            \STATE Update $s_i(t) \leftarrow s_i(t+1)$;
        \ENDFOR
        \FOR{epoch $=1,2,\ldots,E$}
            \STATE Sample a mini-batch from $\mathcal{D}$;
            \STATE Compute advantage using GAE by (\ref{advantage_function_detail});
            \STATE Compute actor clipped surrogate loss by (\ref{PPO_clipp});
            \STATE Update actor network with policy gradient by (\ref{gradient_computed});
            \STATE Compute critic loss by (\ref{MSE});
            \STATE Update critic network with gradient $\nabla_\theta L^V$;
        \ENDFOR
    \ENDFOR
    \end{algorithmic}
\end{algorithm}

\section{Performance Evaluation}\label{sec7}

In this section, we first describe the simulation parameter settings and then comprehensively evaluate the performance of our proposed Ly-DTMPPO algorithm by comparing it against several representative benchmark schemes, and finally discuss the evaluation results.

\subsection{Parameters Setting}\label{subsec7.1}
We adopt the environment of pytorch 1.11 and Python 3.8 with NVIDIA GeForce RTX 4070 SUPER 12GB, 13th Gen Intel(R) Core(TM) i5-13600KF 3.50 GHz, and 32 GB of RAM. The detailed settings are as follows:  

\textit{1) Simulation Scenario:} We consider an east–west oriented four-lane road, where each lane is 3.75 m ($L_{0}$) wide \cite{11023535} and 700 m long. Three RSUs are deployed along the north side of the center road at a distance of 9.375 m ($l_{k}$), with 150 m spacing, 10 m height ($H_{k}$) \cite{9982429}, and 200 m coverage radius. The system bandwidth $\mathcal{B}$ is set to 20 MHz \cite{11023535}, the noise power $\sigma_0^{2}$ is $10^{-13}$ W \cite{9982429}, and the system operates in time slots with a duration of 1 s ($\tau$).  
    
\textit{2) Parameters of CVs and RSUs:} The vehicle speed $v_i^{cv}$ follows a uniform distribution in [12, 16] m/s. The computing capability of each vehicle $C_{CV_i}^{\mathrm{cpu}}$ is uniformly distributed in [2, 3] GHz \cite{11126101}, while each RSU is equipped with a fixed 20 GHz computing capacity ($F_k^{\mathrm{rsu,max}}$). The task data volume $G_{i}^{\mathrm{TasD}}$ is uniformly distributed in [1, 3] Mb \cite{11126101}, and the computation intensity of task $I_{c}^{\mathrm{task}}$ is initialized in the range of [1500, 2000] cycles/bit.

\textit{3) Parameters of training :} For the proposed Ly-DTMPPO algorithm, the learning rate is set to $8\times10^{-5}$, with a discount factor $\gamma=0.99$, GAE parameter $\lambda=0.95$, clipping ratio $\epsilon=0.2$, and 10 update epochs. Each episode consists of 30 time slots, and the maximum number of episodes is 1800. The Adam optimizer is employed for parameter updates to ensure stable convergence.

The values of the remaining parameters are summarized in Table \ref{tab2}.

\begin{table}[htbp]
\caption{Simulation Parameters}
\centering
\setlength{\tabcolsep}{4.5mm}{
\begin{tabular}{p{5cm}|l}
\toprule
\textbf{PARAMETERS} & \textbf{VALUES} \\
\midrule
        Maximum number of episodes: & $1800$ \\
        \hline
        Steps per episode: & $30$\\
        \hline
		Number of Vehicles: $N$  & $[2,6]$ \\ 
        \hline
		Lane width (m): $L_{0}$ & $3.75$\\
        \hline
		Height of RSU (m): $H_{k}$ & $10$\\
        \hline
        Coverage radius of RSU (m): $Range_{k}^{rsu}$ & $200$ \\
        \hline
        Distance between neighboring RSUs (m): & $150$\\
        \hline
        Distance from RSU to center of road (m): $l_{k}$ & $9.75$ \\
		\hline
        Bandwidth (MHz): $\mathcal{B}$ & $20$ \\  
		\hline
        Duration of time slots (s): $\tau$ & $1$ \cite{liu2025lyapunovguideddiffusionbasedreinforcementlearning} \\
		\hline
        Additive white Gaussian noise (W): $\sigma_0^{2}$ & $10^{-13}$ \cite{9982429}\\
		\hline
        Transmission power (W): $P_i$ & $1$  \\ 
		\hline
        Vehicles speed (m/s): $v_i^{cv}$  & $[12,16]$ \\
		\hline
        CV computation capacity (GHz): $C_{CV_i}^{\mathrm{cpu}}$ & $[2,3]$ \cite{10638833}\\
        \hline
        RSU computing resources (GHz): $F_k^{\mathrm{rsu,max}}$ & $20$\\
        \hline
        Task data volume (Mb): $G_{i}^{\mathrm{TasD}}$ & $[1,3]$\\ 
        \hline
        The computational intensity of a task (cycles/bit): $I_{c}^{\mathrm{task}}$ & $[1500,2000]$ \cite{9982429}\\
        \hline
        The computational intensity of the coordinate transformation (cycles/bit): $I_{c}^{\mathrm{co-tra}}$ & $[100,500]$ \cite{9982429}\\
        \hline
        The vehicle energy coefficient: $\kappa_{cv}$ & $10^{-26}$ \cite{9449944}\\
        \hline
        The RSU energy coefficient: $\kappa_{rsu}$ & $10^{-28}$ \cite{9128785}\\
        \hline
        Communication gain factor: $\Theta$ & $1.5$ \cite{10638833}\\
        \hline
        The Lyapunov Control Parameter: $V$ & $5$\\
        \hline
        Batch size: $B$ & $512$ \\
        \hline
        Discount factor: $\gamma $ & $0.99$\\
        \hline
        GAE parameter: $\lambda$ & $0.95$\\
        \hline
        Clipping ratio: $\epsilon$ & $0.2$\\
\bottomrule
\end{tabular}
}
\label{tab2}
\end{table}

\subsection{Baselines}\label{subsec7.2}
To demonstrate the effectiveness of the proposed Ly-DTMPPO algorithm, we compare it with several representative MARL baselines. For fairness, all algorithms are implemented under the same simulation environment and adopt identical parameter settings for the vehicular network scenario described in Sec.~\ref{subsec7.1}. The main differences lie in the decision-making modules for task offloading and resource allocation.

\textit{1) Ly-MADDPG \cite{10115012,10335755}:} This baseline adopts the multi-agent deep deterministic policy gradient (MADDPG) as the backbone, combined with the Lyapunov-based queue stability formulation for resource allocation and task offloading. 

\textit{2) Ly-MATD3:} This baseline replaces the PPO core of the proposed scheme with multi-agent twin delayed deep deterministic policy gradient (MATD3) \cite{10974618}. The Lyapunov optimization framework remains identical, ensuring fair comparison in terms of queue stability. 

\textit{3) Ly-MASAC:} This baseline adopts the multi-agent soft actor-critic (MASAC) algorithm, which is an off-policy algorithm leveraging entropy regularization for improved exploration. It extends the LySAC framework in \cite{10638833} to the multi-agent setting, where the Lyapunov optimization mechanism is incorporated to guarantee queue stability. 

\textit{4) Ly-MAPPO \cite{9895362}:} This baseline applies the MAPPO algorithm under the same Lyapunov optimization framework, but removes the DT-enhanced global visibility and predictive modeling. The comparison with our proposed Ly-DTMPPO highlights the unique contribution of DT integration in improving decision-making efficiency under high-mobility vehicular environments.

\subsection{Simulation Results}\label{subsec7.3}
In this section, we first evaluate the overall system performance of the proposed Ly-DTMPPO algorithm under the default parameter settings. Subsequently, we conduct a comparative analysis with the benchmark schemes to highlight the effectiveness of our approach. Finally, we investigate the influence of key system parameters on the performance of Ly-DTMPPO and the baseline algorithms to provide deeper insights into their adaptability and robustness in dynamic vehicular environments.

\begin{figure}[htbp]
\centering
\includegraphics[width=3.4in]{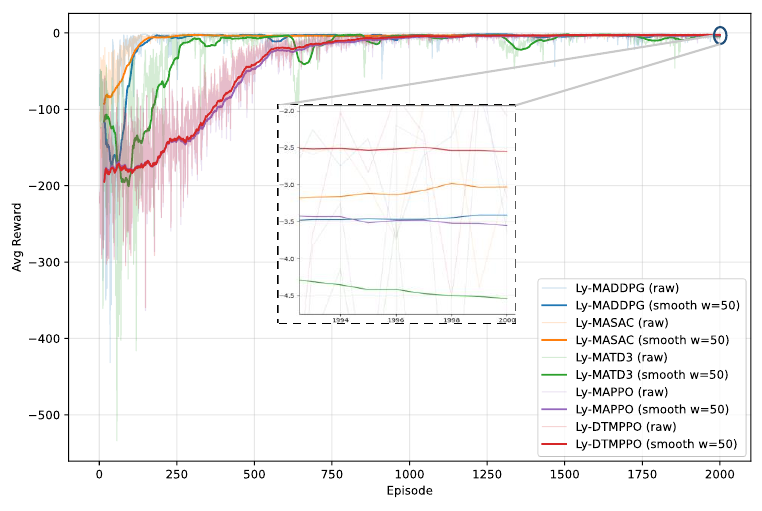}
\caption{Average rewards under different algorithms.} 
\label{Convergence_reward}
\end{figure}

\textit{1) Convergence Performance:} Fig. \ref{Convergence_reward} depicts the convergence behaviors of all schemes. All methods eventually converge, This experiment was conducted with 5 CVs, and 3 RSUs. In comparison, Ly-DTMPPO (red curve) attains the highest steady-state reward and maintains relatively smooth convergence. Using the average reward over the last $50$ episodes as the steady-state metric, Ly-DTMPPO achieves relative gains of 25.3\%, 43.9\%, 15.9\%, and 28.3\% over Ly-MADDPG, Ly-MATD3, Ly-MASAC, and Ly-MAPPO, respectively. Moreover, Ly-DTMPPO exhibits a coefficient of variation (CV) of 0.411, which is significantly smaller than Ly-MAPPO (0.672) and Ly-MADDPG (0.530), and comparable to MASAC (0.366) and MATD3 (0.333), thereby demonstrating superior stability against most baselines. In terms of convergence speed, Ly-DTMPPO reaches 95\% of its steady-state reward at 1711 episodes, which is slower than the off-policy schemes (e.g., Ly-MADDPG at 185 and Ly-MASAC at 278). Nevertheless, the substantially higher reward underscores the advantage of the on-policy PPO backbone combined with DT-enabled global visibility and Lyapunov-based queue stabilization in achieving both optimality and robustness in highly dynamic vehicular networks. The detailed numerical results are further summarized in Table~\ref{tab3}.

\begin{table}[!t]
\centering
\caption{Convergence Stability and Speed of Different Schemes}
\label{tab3}
\begin{threeparttable}
\begin{tabular}{lccc}
\toprule
\textbf{Algorithm} & \textbf{Steady Reward}$^{\mathrm{1}}$ & \textbf{CV}$^{\mathrm{2}}$ & \textbf{Conv. Ep. (95\%)}$^{\mathrm{3}}$ \\
\midrule
Ly-MADDPG  & -3.412 & 0.530 & 185  \\
Ly-MASAC   & -3.031 & 0.366 & 278  \\
Ly-MATD3   & -4.541 & 0.333 & 579  \\
Ly-MAPPO   & -3.555 & 0.672 & 1398 \\
Ly-DTMPPO  & -2.549 & 0.411 & 1711 \\
\bottomrule
\end{tabular}
\begin{tablenotes}
\item $^{\mathrm{1}}$ Steady reward is averaged over the last $50$ episodes. 
\item $^{\mathrm{2}}$ CV denotes the coefficient of variation.
\item $^{\mathrm{3}}$ “Conv. Ep.” represents the first episode achieving 95\% of the steady reward.
\end{tablenotes}
\end{threeparttable}
\end{table}

\begin{figure}[htbp]
\centering
\subfloat[]{
    \includegraphics[width=1.5in]{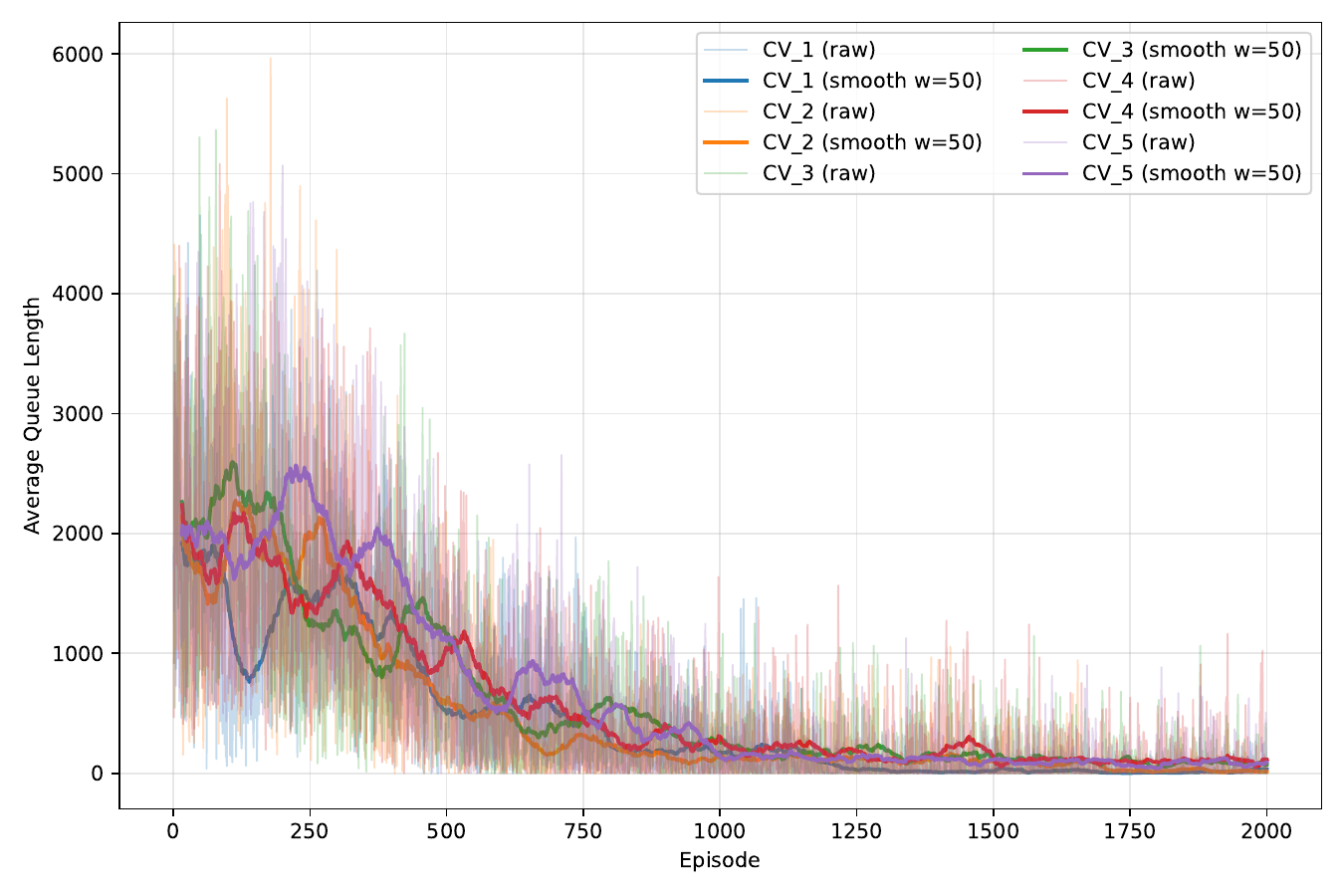}
    \label{Queue_per_CV_LyDTMPPO}
}
\hfil
\subfloat[]{
    \includegraphics[width=1.5in]{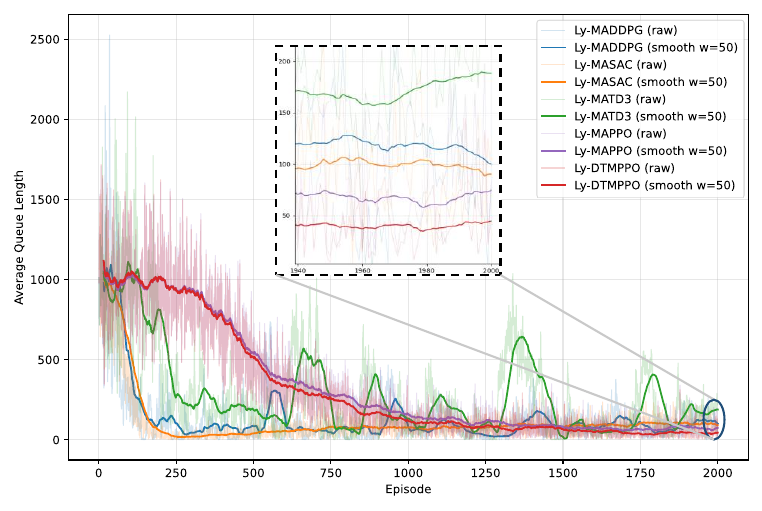}
    \label{Average_Queue_length_comparison}
}
\caption{Task queue length comparison of different CVs and algorithms: (a) Task queue length of different CVs; (b) Task queue length of different algorithms.}
\label{Queue_comparison}
\end{figure}

\textit{2) The Queue Stability Performance:} Fig. \ref{Queue_per_CV_LyDTMPPO} evaluates the queue–stability achieved by integrating Lyapunov optimization into the decision loop. We consider a setting with 5 CVs, 3 RSUs, and Lyapunov control parameter V=5. In Fig. 7(a), the per-vehicle (CV1–CV5) task-queue trajectories rapidly decay from their initial transients to a small steady window and remain bounded thereafter, evidencing that the Lyapunov drift terms effectively suppress backlog growth even under heterogeneous arrivals. To make the trend visually clear, the curves are shown both in raw form and with a moving-average smoothing over the last 50 episodes; both views consistently indicate fast convergence and low residual oscillation across all agents, which implies stable task admission/offloading and robust queue regulation at the vehicle side.

Fig. \ref{Average_Queue_length_comparison} compares the average system queue length (averaged over all CVs per episode) across algorithms. Compared with other algorithms, our Ly-DTMPPO curve exhibits (i) a steeper descent phase (faster backlog clearance) and (ii) a lower steady-state plateau with reduced fluctuation. This behavior aligns with the design goals: the drift-plus-penalty mechanism enforces queue stability, while the DT-enhanced global visibility improves prediction of workload and channel dynamics, enabling more queue-aware resource allocation. Overall, the convergence of the time-evolving queues demonstrates that the proposed scheme leads to a stable task-assignment process and bounded queues, thereby guaranteeing stable system operation in the considered DT-enabled ISAC-aided IoV scenario.

\begin{figure*}[htbp]
\centering
\subfloat[]{
    \includegraphics[width=2.2in]{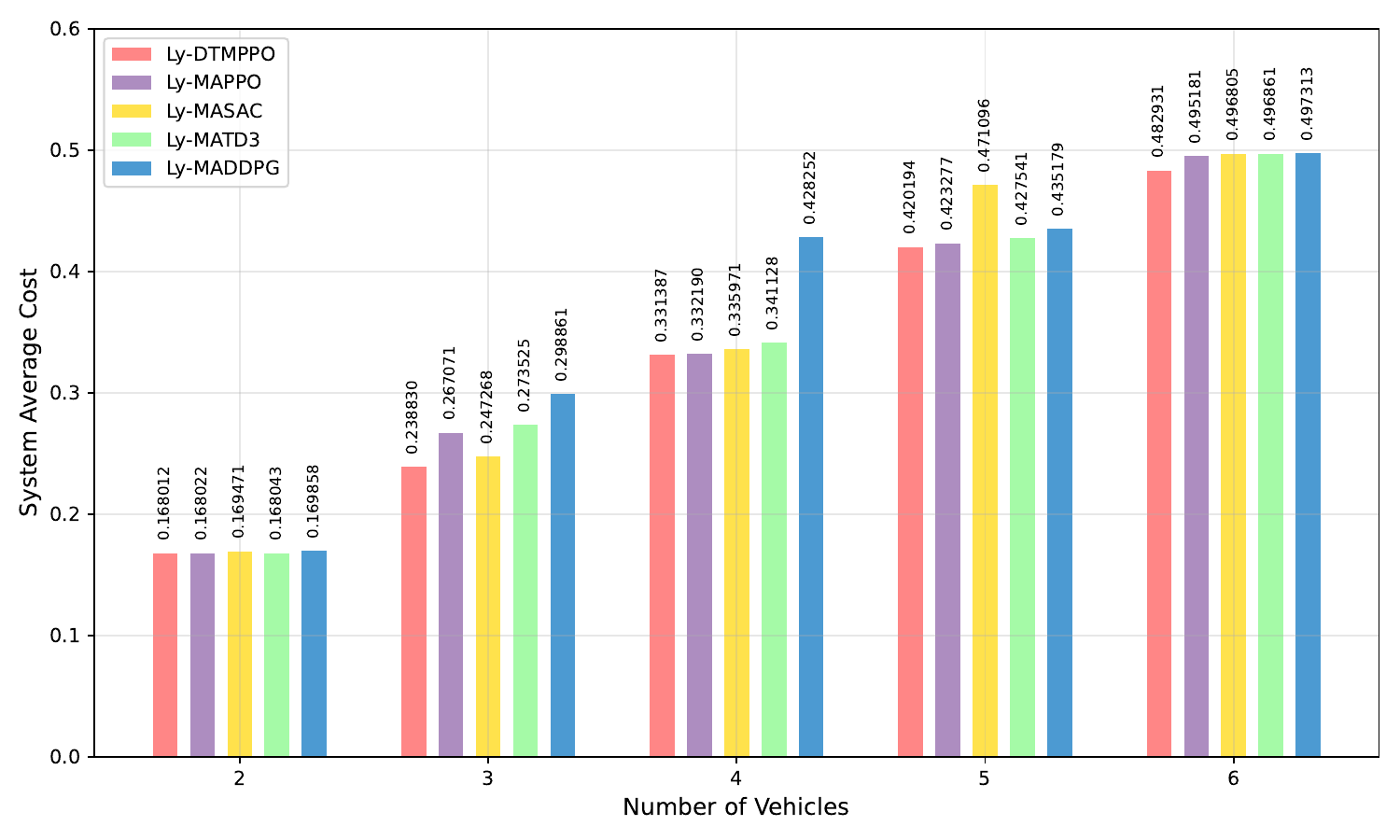}
    \label{sys_cost_different_CV}
}
\hfil
\subfloat[]{
    \includegraphics[width=2.2in]{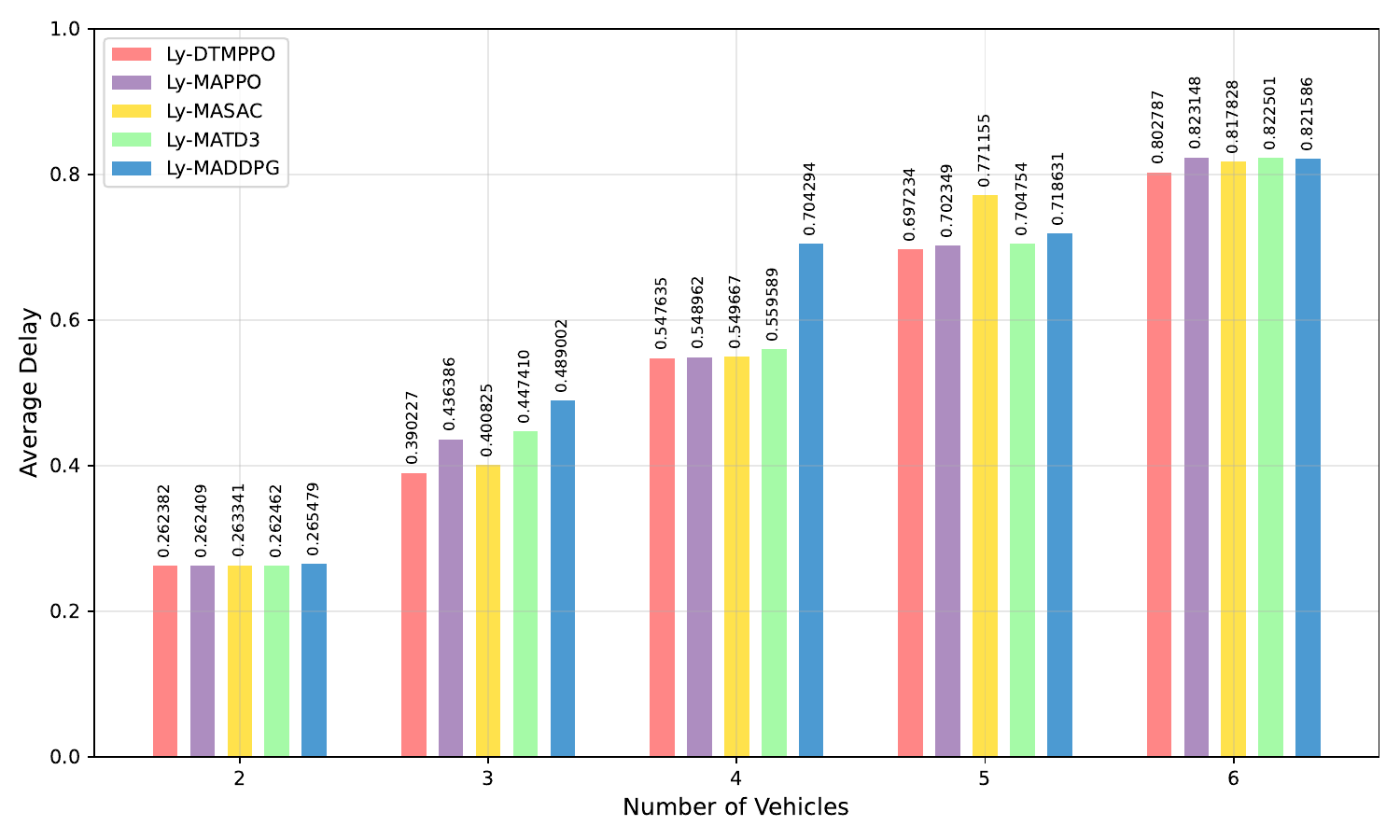}
    \label{sys_delay_different_CV}
}
\hfil
\subfloat[]{
    \includegraphics[width=2.2in]{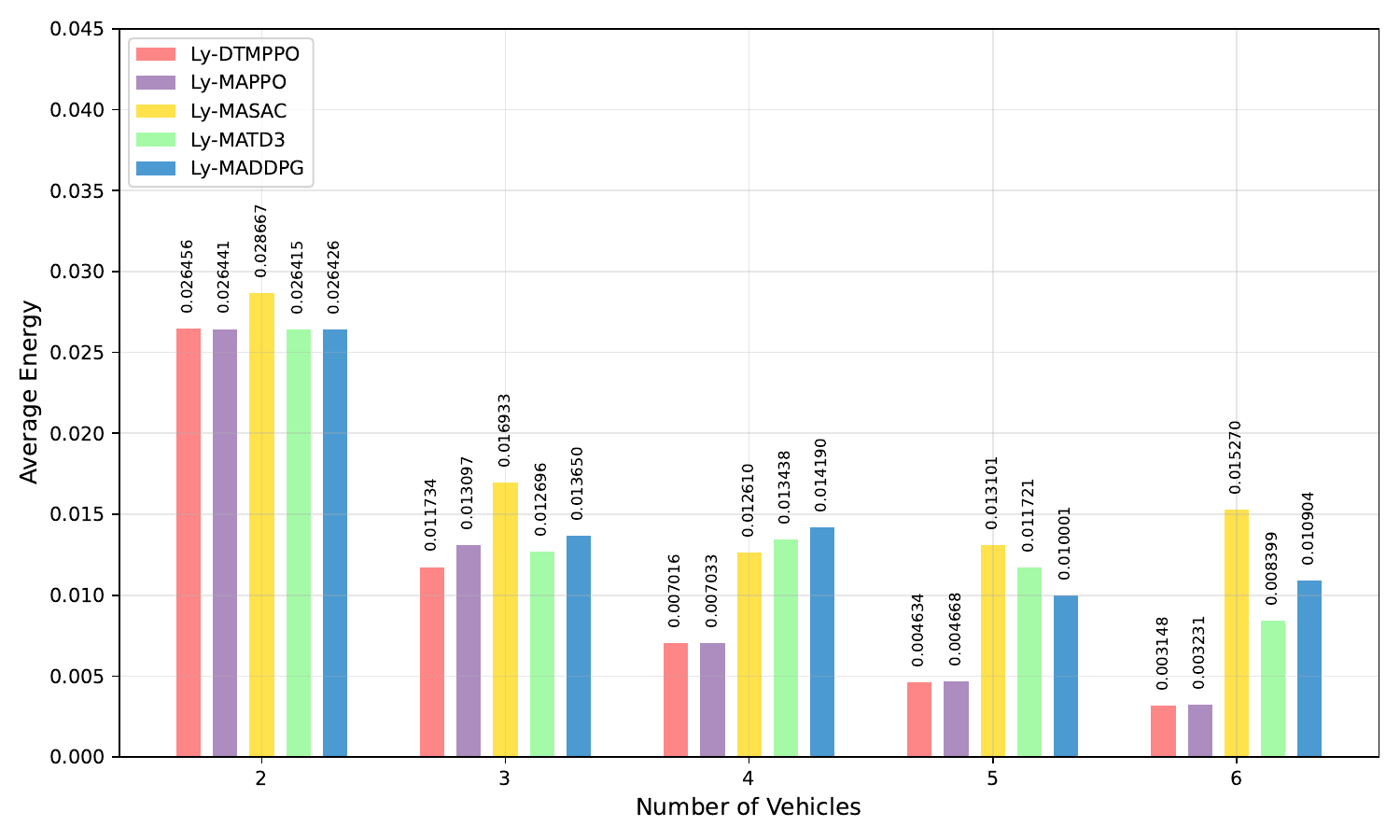}
    \label{sys_energy_different_CV}
}
\caption{System performance versus the number of CVs under different algorithms: (a) System average cost versus the number of CVs under different algorithms; (b) Average delay versus the number of CVs under different algorithms; (c) Average energy consumption versus the number of CVs under different algorithms.}
\label{sys_performance_vs_CV}
\end{figure*}

\textit{3) Effect of the Number of CVs:} Fig. \ref{sys_cost_different_CV} shows the effect of the number of CVs on the system average cost. This comparison is conducted under the condition where the configurable weight $\alpha$ is set to 0.6, indicating the system prioritizes system delay more than energy consumption. We can observe that the system average cost consistently increases with the number of CVs for all compared algorithms, due to the higher computational and communication load brought by more CVs. However, across all traffic loads (2–6 CVs), the proposed Ly-DTMPPO consistently achieves the lowest cost. Taking two representative loads as examples, with 3 CVs Ly-DTMPPO reduces the cost by 10.57\%, 3.41\%, 2.68\%, and 20.09\% compared with Ly-MAPPO, Ly-MASAC, Ly-MATD3, Ly-MADDPG, respectively. At 4 CVs, the reductions are 0.24\%, 1.36\%, 2.86\%, and 22.62\%, respectively. Averaged over all loads (2–6 CVs), Ly-DTMPPO yields 2.81\%, 3.85\%, 4.02\%, and 10.03\% lower cost than Ly-MAPPO, Ly-MASAC, Ly-MATD3, and Ly-MADDPG, respectively. This demonstrates that incorporating Lyapunov-guided optimization and DT decision-making effectively balances delay and energy overheads, leading to superior overall performance.

Fig. \ref{sys_delay_different_CV} presents the average delay performance under different numbers of CVs. As expected, the delay consistently increases with the number of vehicles, due to the heavier traffic load and competition for limited communication and computing resources. Among all schemes, Ly-DTMPPO maintains the lowest average delay across all scenarios. For example, at 3 CVs, Ly-DTMPPO reduces the delay by 10.6\%, 20.2\%, 2.6\%, and 12.8\% compared with Ly-MAPPO, Ly-MADDPG, Ly-MASAC, and Ly-MATD3, respectively. On average (2-6 CVs), Ly-DTMPPO achieves delay reductions of approximately 2.8\%, 9.8\%, 3.0\%, and 3.7\% relative to these baselines. These results demonstrate that the combination of PPO’s on-policy exploration with DT-enhanced global visibility and Lyapunov stabilization yields more efficient resource allocation, thereby alleviating queue accumulation and lowering the overall task processing latency.

Fig. \ref{sys_energy_different_CV} presents the average system energy consumption under different numbers of CVs. It can be observed that Ly-DTMPPO and Ly-MAPPO exhibit a monotonic decrease from 2→6 CVs, indicating progressively improved coordination between local computing and RSU offloading. In contrast, Ly-MASAC shows a decrease up to 4 CVs but rebounds at 5–6 CVs; Ly-MATD3 increases at 4 CVs and then drops; Ly-MADDPG fluctuates notably. These non-monotonic behaviors reveal that off-policy baselines may induce unbalanced RSU utilization and redundant offloading under high mobility, leading to oscillatory energy consumption. 

For example, at 6 CVs,  Ly-DTMPPO reduces energy consumption by 2.6\%, 79.4\%, 62.5\%, and 71.1\% compared to Ly-MAPPO, Ly-MASAC, Ly-MATD3, and Ly-MADDPG, respectively. Overall, Ly-DTMPPO achieves an average energy consumption improvement of approximately 2-3\% compared to Ly-MAPPO across the entire load range (2-6 CVs), while the average improvements compared to other schemes are approximately 46\%, 36\%, and 37\%, respectively. This indicates that, under the combined effect of Lyapunov optimization and DT global visibility, Ly-DTMPPO can effectively avoid invalid local computations and RSU overload, thereby significantly reducing the total system energy consumption while ensuring quality of service.

\begin{figure}[htbp]
\centering
\subfloat[]{
    \includegraphics[width=1.63in]{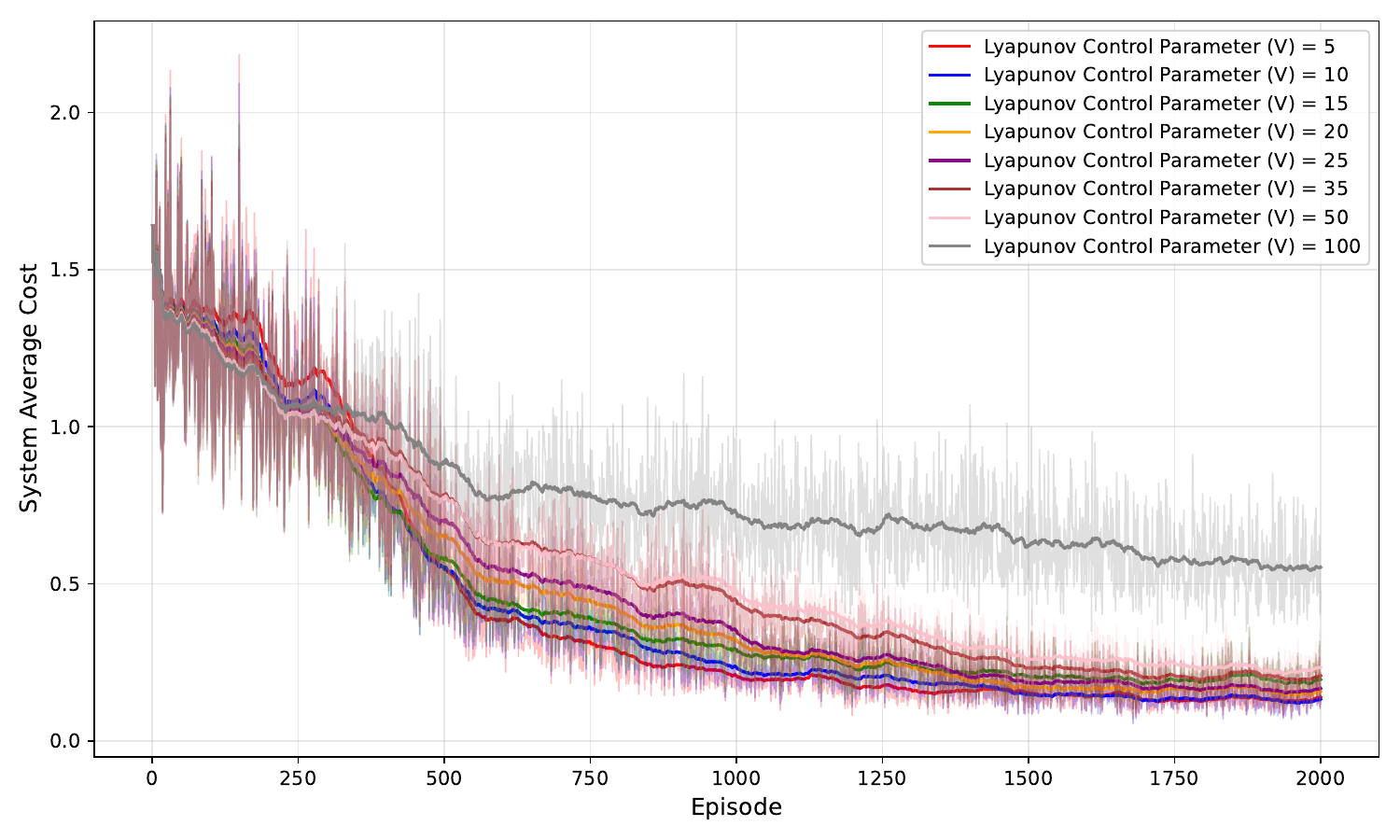}
    \label{sys_cost_VS_Episode_different_V}
}
\hfil
\subfloat[]{
    \includegraphics[width=1.63in]{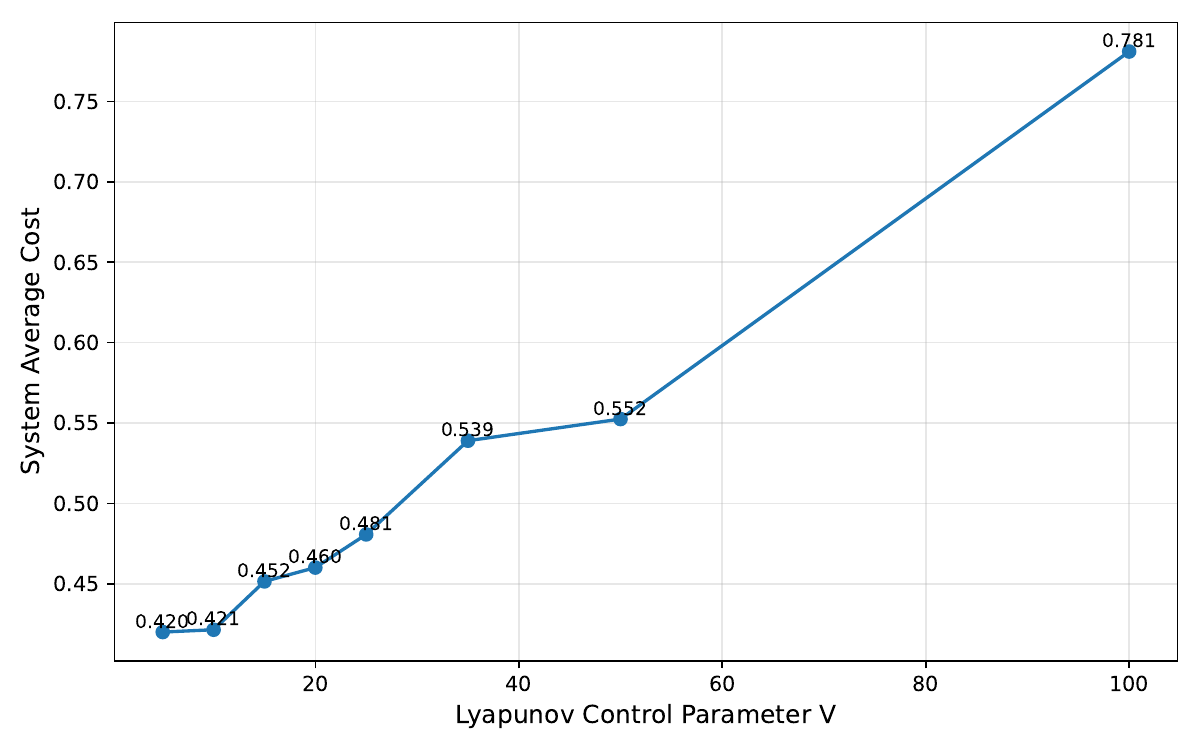}
    \label{sys_cost_different_V}
}
\caption{System performance comparison under different Lyapunov control parameters: (a) System average cost versus Episode under different Lyapunov \\ control parameter $V$; (b) System average cost versus Lyapunov control parameter $V$.}
\label{sys_cost_Lyapunov_comparison}
\end{figure}

\textit{4) Effect of the Lyapunov Control Parameter V:} Fig.~\ref{sys_cost_VS_Episode_different_V} and Fig.~\ref{sys_cost_different_V} illustrate the impact of the Lyapunov control parameter $V$ on the system average cost. As shown in Fig.~\ref{sys_cost_VS_Episode_different_V}, all settings of $V$ eventually lead to convergence. However, the steady-state levels differ significantly. For small $V$ (e.g., $V=5$ or $10$), the system achieves relatively low average costs (around 0.42), since the drift-penalty tradeoff emphasizes queue stability, thereby suppressing excessive task backlog. As $V$ increases, the average cost gradually rises (e.g., reaching 0.552 at $V=50$ and 0.781 at $V=100$), due to larger average backlog accumulation across task queues, which amplifies the delay component in the overall cost.

This trend indicates that an excessively large $V$ prioritizes immediate energy minimization but neglects delay control, while an overly small $V$ may lead to aggressive queue draining at the expense of energy efficiency. Therefore, selecting a moderate $V$ achieves the most balanced tradeoff, ensuring minimal system average cost while preserving Lyapunov stability. These results highlight the importance of properly tuning the Lyapunov control parameter to fully exploit the drift-penalty framework in dynamic vehicular environments.
\section{Conclusion}\label{sec8}
In this paper, we have studied DT-assisted task offloading and resource allocation in lSAC-enabled loV. By leveraging Lyapunov optimization, we have transformed the long-term stochastic control problem into per-slot problems that guarantee queue stability while balancing latency and energy consumption. To further exploit the global visibility of DT, we have proposed the Ly-DTMPPO algorithm, which embeds Lyapunov-based stability modeling into a CTDE multi-agent reinforcement learning framework. This design enables each vehicle to make adaptive and stable decisions using global context information provided by the DT space. Simulation results have validated that Ly-DTMPPO achieves faster convergence, lower delay, and improved energy efficiency compared with existing Lyapunov-driven and multi-agent baselines, resulting in the lowest overall system cost. These findings demonstrate that integrating Lyapunov optimization with DT-enhanced policy learning provides a robust and scalable framework for intelligent resource management in ISAC-enabled IoV. Future work will extend the proposed framework to heterogeneous vehicular edge networks and explore GenAI-based robust optimization techniques \cite{zhao2025generative} to enhance system adaptability under dynamic vehicular and spectrum conditions. 


{\appendices
\section*{Proof of LEMMA 1}
\vspace{-2.65mm}
To begin with, by squaring both sides of (\ref{Local_queue}), (\ref{RSU_queue}), and (\ref{virtual_queues}), we have 
\vspace{-2.6mm}
\begin{align}
& Q_{i}^{\mathrm{loc}}(t+1)^{2}=\begin{bmatrix} 
\mathrm{max}\{Q_{i}^{\mathrm{loc}}(t)-{C_{CV_i}^{\mathrm{cpu}}\tau}+ {\lambda _{i}^{\mathrm{loc}}(t)},0\}
\end{bmatrix}^2, \notag \\
& \forall \mathcal{C}_i\in \mathbb{C}, \label{squaring_Qiloc}\\
& Q_{k}^{\mathrm{rsu}}(t+1)^{2} = [ \mathrm{max}\{Q_{k}^{\mathrm{rsu}}(t)-F_{RSU_k}^{\mathrm{cpu}}(t)\tau \notag \\
& + \displaystyle \sum_{ \mathcal{C}_i \in \mathbb{C}} \lambda _{i}^{\mathrm{rsu}}(t), 0\}]^2, \forall \mathcal{R}_k\in \mathbb{R}, \label{squaring_Qkrsu} \\
& V_{i}(t+1)^{2}=\begin{bmatrix}
\mathrm{max}\{V_{i}(t) - E^{\mathrm{max}} + E^{\mathrm{total}}(t),0\} 
\end{bmatrix}^2. \notag \\
& \forall \mathcal{C}_i\in \mathbb{C}. \label{squaring_Vi}
\end{align}
\vspace{-6mm}

Then, we consider inequality $[\max\{x,0\}]^2 \le x^2$, (\ref{squaring_Qiloc}), (\ref{squaring_Qkrsu}), and (\ref{squaring_Vi}) can be transformed into the following equations, respectively.
\vspace{-1mm}
\begin{align}
& Q_{i}^{\mathrm{loc}}(t+1)^{2} - Q_{i}^{\mathrm{loc}}(t)^{2} \le  2 Q_{i}^{\mathrm{loc}}(t)({\lambda _{i}^{\mathrm{loc}}(t)} - {C_{CV_i}^{\mathrm{cpu}}\tau}) \notag \\
& + ({\lambda _{i}^{\mathrm{loc}}(t)} - {C_{CV_i}^{\mathrm{cpu}}\tau})^2, \label{inequality1_Qiloc} \\
& Q_{k}^{\mathrm{rsu}}(t+1)^{2} - Q_{k}^{\mathrm{rsu}}(t)^{2} \le 2Q_{k}^{\mathrm{rsu}}(t) (\sum_{\mathcal{C}_i\in \mathbb{C}}\!\! \lambda _{i}^{\mathrm{rsu}}(t) \!-\! F_{RSU_k}^{\mathrm{cpu}}(t)\tau) \notag \\
& + (\sum_{\mathcal{C}_i\in \mathbb{C}} \lambda _{i}^{\mathrm{rsu}}(t) - F_{RSU_k}^{\mathrm{cpu}}(t)\tau)^2, \label{inequality1_Qkrsu}\\
& V_{i}(t+1)^{2} - V_{i}(t)^{2} \le 2 V_{i}(t) (E^{\mathrm{total}}(t) - E^{\mathrm{max}}) \notag \\
& + (E^{\mathrm{total}}(t) - E^{\mathrm{max}})^2. \label{inequality1_Vi}
\end{align}
\vspace{-5mm}

Next, summing (\ref{inequality1_Qiloc}) over all CVs. Similarly, (\ref{inequality1_Qkrsu}), and (\ref{inequality1_Vi}) have the same operator. They hold:
\vspace{-2mm}
\begin{align}
& \frac{1}{2} \sum_{\mathcal{C}_i\in \mathbb{C}} [Q_{i}^{\mathrm{loc}}(t+1)^{2} - Q_{i}^{\mathrm{loc}}(t)^{2}] \le \frac{1}{2}\sum_{\mathcal{C}_i\in \mathbb{C}} \notag \\
& \Bigg[ 2 Q_{i}^{\mathrm{loc}}(t)({\lambda _{i}^{\mathrm{loc}}(t)} - {C_{CV_i}^{\mathrm{cpu}}\tau}) + ({\lambda _{i}^{\mathrm{loc}}(t)} - {C_{CV_i}^{\mathrm{cpu}}\tau})^2\Bigg ], \label{inequality2_Qiloc}\\
& \frac{1}{2} \sum_{\mathcal{R}_k \in \mathbb{R}} [Q_{k}^{\mathrm{rsu}}(t+1)^{2} - Q_{k}^{\mathrm{rsu}}(t)^{2}] \le \frac{1}{2} \sum_{\mathcal{R}_K\in \mathbb{R}} \notag \\ 
&  \Bigg [ 2 Q_{k}^{\mathrm{rsu}}(t) (\sum_{\mathcal{C}_i\in \mathbb{C}} \lambda _{i}^{\mathrm{rsu}}(t) - F_{RSU_k}^{\mathrm{cpu}}(t)\tau) \notag \\
& + (\sum_{\mathcal{C}_i\in \mathbb{C}} \lambda _{i}^{\mathrm{rsu}}(t) - F_{RSU_k}^{\mathrm{cpu}}(t)\tau)^2 \Bigg ], \label{inequality2_Qkrsu}\\
& \frac{1}{2} \sum_{\mathcal{C}_i\in \mathbb{C}} [V_{i}(t+1)^{2} - V_{i}(t)^{2}] \le \frac{1}{2} \sum_{\mathcal{C}_i\in \mathbb{C}} \notag \\
& \Bigg [ 2 V_{i}(t) (E^{\mathrm{total}}(t) - E^{\mathrm{max}}) + (E^{\mathrm{total}}(t) - E^{\mathrm{max}})^2 \Bigg ]. \label{inequality2_Vi}
\end{align}
We define
\vspace{-3mm}
\begin{align}
& L(V_i(t))=\frac{1}{2} \sum_{\mathcal{C}_i\in \mathbb{C}}V_{i}(t)^{2}, \\
& \Delta(V_i(t)) = \mathbb{E} \left[ L(V_i(t+1)) - L(V_i(t)) \mid \mathbf{Z}(t) \right].
\end{align}
And by taking the conditional expectation on both sides of (\ref{inequality2_Qiloc}), (\ref{inequality2_Qkrsu}), and (\ref{inequality2_Vi}), we have
\begin{align}
& \Delta(Q_{i}^{\mathrm{loc}}(t)) \le B_1(t) + \! \sum_{\mathcal{C}_i\in \mathbb{C}} \!\!\mathbb{E} \left[ Q_{i}^{\mathrm{loc}}(t)({\lambda _{i}^{\mathrm{loc}}(t)} \!-\! {C_{CV_i}^{\mathrm{cpu}}\tau}) \! \mid \! \mathbf{Z}(t) \right], \notag \\
& B_1(t) = \frac{1}{2} \sum_{\mathcal{C}_i\in \mathbb{C}}\mathbb{E} \left[ ({\lambda _{i}^{\mathrm{loc}}(t)} - {C_{CV_i}^{\mathrm{cpu}}\tau})^2 \mid \mathbf{Z}(t) \right], \label{inequality3_Qiloc} \\
& \Delta(Q_{k}^{\mathrm{rsu}}(t)) \le B_2(t) \notag \\
& +  \sum_{\mathcal{R}_k\in \mathbb{R}} \mathbb{E} \left[ Q_{k}^{\mathrm{rsu}}(t) (\sum_{\mathcal{C}_i\in \mathbb{C}} \lambda _{i}^{\mathrm{rsu}}(t) - F_{RSU_k}^{\mathrm{cpu}}(t)\tau)
\mid \mathbf{Z}(t) \right], \notag \\
& B_2(t) = \frac{1}{2} \sum_{\mathcal{R}_k\in \mathbb{R}}\mathbb{E} \left[ (\sum_{\mathcal{C}_i\in \mathbb{C}} \lambda _{i}^{\mathrm{rsu}}(t) - F_{RSU_k}^{\mathrm{cpu}}(t)\tau)^2 \mid \mathbf{Z}(t) \right], \label{inequality3_Qkrsu}\\
& \Delta(V_i(t)) \le B_3(t) +  \sum_{\mathcal{C}_i\in \mathbb{C}} \mathbb{E} \left[ V_{i}(t) (E^{\mathrm{total}}(t) - E^{\mathrm{max}}) \mid \mathbf{Z}(t) \right], \notag \\
& B_3(t) = \frac{1}{2} \sum_{\mathcal{C}_i\in \mathbb{C}} \mathbb{E} \left[ (E^{\mathrm{total}}(t) - E^{\mathrm{max}})^2 \mid \mathbf{Z}(t) \right]. \label{inequality3_Vi}
\end{align}
\vspace{-3mm}
Summing over the four inequalities in (\ref{inequality3_Qiloc}), (\ref{inequality3_Qkrsu}), and (\ref{inequality3_Vi}):
\begin{align}
& \Delta(\mathbf{Z}(t)) = \mathbb{E} \bigg[ \frac{1}{2} \sum_{\mathcal{C}_i\in \mathbb{C}} \left( Q_{i}^{\mathrm{loc}}(t+1)^2 - Q_{i}^{\mathrm{loc}}(t)^2 \right) \notag \\
& + \frac{1}{2} \sum_{\mathcal{R}_k \in \mathbb{R}} \left( Q_{k}^{\mathrm{rsu}}(t+1)^2 - Q_{k}^{\mathrm{rsu}}(t)^2 \right) \notag \\
& + \frac{1}{2} \sum_{\mathcal{C}_i\in \mathbb{C}} \left( V_{i}(t+1)^2 - V_{i}(t)^2 \right) 
\ \bigg| \ \mathbf{Z}(t) \bigg] \notag \\
& \le B(t) + \mathbb{E} \bigg[ \sum_{\mathcal{C}_i\in \mathbb{C}} Q_{i}^{\mathrm{loc}}(t)({\lambda _{i}^{\mathrm{loc}}(t)} - {C_{CV_i}^{\mathrm{cpu}}\tau}) \notag \\
& + \sum_{\mathcal{R}_k\in \mathbb{R}} Q_{k}^{\mathrm{rsu}}(t) (\sum_{\mathcal{C}_i\in \mathbb{C}} \lambda _{i}^{\mathrm{rsu}}(t) - F_{RSU_k}^{\mathrm{cpu}}(t)\tau) \notag \\
& + \sum_{\mathcal{C}_i\in \mathbb{C}} V_{i}(t) (E^{\mathrm{total}}(t) - E^{\mathrm{max}})
\ \bigg| \ \mathbf{Z}(t) \bigg],
\end{align}
where $B(t)=B_1(t)+B_2(t)+B_3(t)$. 

Finally,  by grouping the elements in the upper bound that are independent of the queue states into a constant term $B$, and by determining the offloading decisions $\ell(t)$, the bandwidth allocation $\textbf{b}(t)$, the computation resource allocation $F(t)$, and the energy consumption term $E^{total}(t)$, all $B_1(t)$, $B_2(t)$, $B_3(t)$ become deterministic constants at each time slot. Thus, the overall upper bound $B(t)$ reduces to a constant $B$, which completes the proof of Lemma 1.}



\bibliographystyle{IEEEtran}

\newpage

\vfill

\end{document}